# How does hazard exposure influence job choice? Evaluating time-dependent tradeoffs between salary and hazard risks

Richard Bernknopf [1*], Leila Gonzales [1], and Christopher Keane [1]

July 15, 2026

## Abstract

Natural hazards are a nonmarket disamenity that affects an individual's search for employment resulting in a negative environmental impact that produces an economic inefficiency. We develop a seek-and-screen job search approach that uses a discrete choice simulation to examine how salary, crime, and natural hazard risk influence job choice. We model the job decision process as a series of elimination events using a Cox hazard model grounded in a Random Utility Model. We use data from the discrete choice simulation to estimate both a standard proportional hazards model and an extended specification that allows the effect of natural hazard risk to vary across decision rounds. Individuals are exposed to the dynamics of a simulated job search as they make decisions between pairs of job offers in an adaptive learning process based on income, geography, crime level, and natural hazard attributes. The results of the job choice decisions provide the input to a statistical survival analysis. The results indicate that salary and crime exert stable and economically intuitive effects on job elimination, with higher salary reducing and higher crime increasing the likelihood of removal. In contrast, natural hazard risk exhibits a time-varying effect that increases the probability of elimination in early rounds but becomes neutral or favorable in later stages of the decision process. These findings suggest that environmental risk is evaluated differently as individuals transition from initial screening to final job selection, highlighting the importance of modeling job choice as a multi-stage process.

[1] American Geosciences Institute, 4220 King Street, Alexandria, VA, 22302, USA
* Corresponding author email: rbernknopf@americangeosciences.org.
This paper is part of the American Geosciences Institute's GRANDE project which was supported by the National Science Foundation under Grant #2223004. Any opinions, findings, and conclusions or recommendations expressed in this material are those of the author(s) and do not necessarily reflect the views of the National Science Foundation. We thank the following individuals for their comments on prior drafts of the paper: David Brookshire, Yusuke Kuwayama, and Matthew Sloggy.

## 1. Introduction

Natural hazards impact society as nonmarket spillovers that affect individuals, economic sectors and public institutions, producing economic inefficiencies in the form of a negative societal amenity. The effect of environmental spillovers can result in disaster costs that have long term impacts on people and property (Fajgelbaum and Gaubert 2020, H. John Heinz III Center for Science, Economics and the Environment 2000). In January 2025, the Los Angeles County wildfire disaster included several incidents including the Pacific Palisades, Eaton, Lidia, Kenneth, Hurst, Hughes Fires and lesser fires that burned about 57,529 acres or 90 miles$^2$ and destroyed or damaged at least 18,189 structures causing an estimated total economic cost of $250 Billion (Los Angeles Times 2026, California Department of Forestry and Fire Protection 2025). The 2025 Los Angeles, CA wildfire disaster is a prime example of how increasing disaster severity from natural hazards spill over into insurance and real estate markets and create community and socioeconomic vulnerabilities (Gourevitch et al 2023). While natural hazards vary widely over severity levels, geography and recurrence frequency, most individuals do not consider natural hazard impacts as much as income or crime when conducting a job search (Kennan and Walker 2011).

This paper is about job choice and labor market behavior. We integrate a Random Utility Model (RUM) framework with a sequential hazard-based representation of decision-making. Prior research has included discrete choice models of job selection as a single-stage optimization problem. Instead, we demonstrate how the RUM can be represented as a multi-stage screening process in which alternatives are sequentially eliminated. Capitalizing on the equivalence between the Cox partial likelihood (Han and Hausman 1990) and the connection of discrete choice logit probabilities with the Cox Proportional Hazard (PH) model over the risk set (Allison, 1982), we provide a dynamic interpretation of job choice that allows preferences to be revealed differently across stages of the decision process.

The analysis contributes to the role of environmental and geospatial risks in labor market outcomes. Existing studies typically treat natural hazard exposure as a static disamenity, assuming that its effect on job choice is constant. In contrast, our results demonstrate that the influence of natural hazard risk is non-monotonic and stage-dependent. Specifically, hazard risk acts as a negative screening criterion in early rounds but becomes neutral, and then favorable among the final set of alternatives. This finding suggests that individuals initially avoid riskier locations when the choice set is large but are willing to accept such risks when they are compensated by other job attributes or when the set of viable alternatives narrows.

Our methodology shows that an Extended Cox Hazard (ECH) model provides a framework for capturing non-proportional and time-varying effects in discrete choice settings. By allowing the marginal effect of hazard risk to evolve over multiple rounds, the model uncovers patterns that would be obscured in a standard proportional hazards or static discrete choice specification. The approach illustrates how survival analysis can be used to model complex, multi-stage decision processes.

To implement the model, we construct a choice experiment to elicit individual preferences toward crime and natural hazards. Using Random Utility Theory (RUT) allows us to incorporate the heterogeneity among individuals who seek jobs and how these individuals weigh local disamenities of crime (personal) and disasters (nature-based). The choice experiment is a sequential search process that focuses on individuals' perceptions, responses, and insights about reducing or eliminating natural hazards.

Based on RUT, a job offer is defined as a portfolio of factors in which the specific attributes are randomly populated. We pre-define a set of 16 job profiles that specify the allowable salary-to-cost of living ratio, crime risk level, and natural hazard risk level. Each of the job offers in the set presented to the participant is tied to a specific job profile through a job ID (i.e. 1 to 16). The set of job offers is created through a process that first filters possible locations based on the participant's job search locations and then filters those locations to only include locations that meet the eight combinations of crime risk and natural hazard risks defined by the job profiles. For example, given a participant who has specified that they wish to search for jobs in the Pacific and South Atlantic US Census Bureau divisions, all locations within those divisions would be selected for the initial set of job locations. From that set, Tacoma, Washington and Conway, South Carolina both have relatively high natural hazard risk (4), and high crime risk (4) which would make these suitable locations for job profile IDs 1, 11, 12, and 13. These two cities would be selected along with all other cities that meet the crime and natural hazard risk conditions of the job profile criteria to create a pool of possible locations for each job profile. In the example given, Tacoma, Washington and Conway, South Carolina would appear as possible locations for job profile IDs 1, 11, 12, and 13.

For each job profile ID, a salary is calculated for each possible location based on a random selection of a multiplier from the job profile ID's salary-to-cost of living ratio (i.e. for job Profile ID 1, a number between 1.21 and 1.5) and the location's cost of living. For example, the annual cost of living in Tacoma, Washington is $48,744. In this example, randomly selected multipliers for job profile IDs 1, 11, 12, and 13 are: 1.4, 1, 1.25, and 3.5, respectively. This yields the following salaries for profile IDs 1, 11, 12, and 13: $68,242, $48,744, $60,930, $170,604. At the end of this step, each job profile ID will have a set of possible job locations, each with a job salary. Next, for each job profile ID, one job location is randomly selected. This creates the first layer of each job offer (location, crime risk, natural hazard risk, salary).

Next, the job details layer is created. The participant provided information about the highest educational level attained and associated degree field is used to filter job details (title, description, employer name, location, salary range) from the pool of all job details. For each job profile ID's job offer, all jobs that match the US state of the job offer's location and where the job offer's salary is equal to or exceeds the 10th percentile salary are selected as a potential job detail candidate for the job profile ID. From this pool of potential candidates, one job detail candidate is randomly selected as the set of details associated with the job offer. This step completes the construction of the set of 16 job offers presented to the participant.

To accommodate for the temporal aspect of the analysis, primary data are collected in a simulated search of job offers over four sequential rounds $\left(\mathrm{R}(n)\right)$, where $n = 0,1,2,3,4$. A hazard is the probability a job offer is eliminated in a specific round because it contains undesirable attributes. Job choice is a sequential elimination process derived from a RUM. Applying a sequence of conditional logit choice probabilities over the hazard risks, we include time-varying preferences, application to geospatial job attributes, and integration of hazard risk into labor choice. The dynamic process is a sorting of jobs to identify the longest surviving or best offer that becomes the final job choice ($FJC$). During each round, a participant in the choice experiment acquires data and updates their information concerning income, crime levels and natural hazards severity. Observation of participant choices suggests natural hazard risks gain traction as a latent accumulation of information in the process that they use in later rounds of the experiment. We use the results of the experiment to assess an anticipated affordability of long run natural hazard risk protection. To our knowledge, this is among the first studies to document a sign reversal in the effect of environmental risk over the course of a sequential job choice process.

The remainder of the paper is divided into seven sections. First, there is a brief review of the existing literature related to labor force participation modeling and the economics of natural hazards risk. We focus on the review of theoretical and empirical economics market and nonmarket models. We assume that demanders (employers) and suppliers (employees) collectively prefer to minimize hazard risk. In the next section, we introduce the RUM to evaluate individual job attributes as a sequential elimination process of hazard risks. The ECH model is developed for capturing non-proportional and time-varying effects in a discrete choice setting. In the following section, the choice experiment is described. The modeling framework uses the data collected in the experiment and from secondary public data sources to simulate a prospective job search. In the process, participants become aware of the tradeoffs between economic benefits and nonmarket disamenities of a job location. We refer you to Gonzales et al (2026) for details on the components involved in the design of and process of primary data collection for the multiperiod choice experiment. The following section contains the empirical results of the experiment identified by an individual who either accepts or rejects a particular set of income and risk levels associated with a job offer. The approach involves active, dynamic choices based on logic, reason, and scientific information for risk assessment and management, and experiential risk as feelings that are fast, instinctive, and intuitive reactions to risk and salary information necessary to capture the analytic and experiential aspects of risk taking (Slovic et al 2004). The learning process continues with updated offers and eventually terminates with the preferred offer option. Rejected offers are based on eliminating attributes that lose appeal as the individual makes sequential choices in the job search (Kiefer 1988). The obvious outcome of the simulations is that high salary jobs that have low financial burden associated with risk protection are preferred compared to all jobs offered. Interestingly, high salary jobs also can have an outcome that reveals the acceptability of a high financial burden of protection. Next, we use the seek-and-screen process to develop an individual's anticipated affordability ($AA$) to protect themselves from the natural hazard disamenity. We introduce a natural hazard financial burden ratio ($BR$) for an individual. $BR$ is based on disposable income of the $FJC$ and the prevention and preparedness cost that the individual would be willing to pay to avoid damage from a disaster. A discussion follows about the application of the model and the importance of a sign flip on the natural hazards risk effect. Finally, we provide some closing thoughts about how a model like this one can assist individuals to minimize their exposure to hazard risk and to inform them about the financial burden of loss avoidance.

## 2. Background

Hazard risks are a disamenity associated with a job choice that can be evaluated as a component in a model of an individual looking for a job (Lancaster 1997). The model is a dichotomous choice to either accept or reject a job offer based on a group of time dependent and time independent covariates. The outcome of the process contains information that can assist individuals in making choices in investments in natural hazard loss avoidance. We utilize a RUM convolved with a Cox hazard model to identify the correlation of job acceptance and salary, disposable income, crime level and natural hazard severity. Analysis involves a reference case, where job seekers ignore or disregard hazard risk information in their employment decision. In this case, geographic variability is not a factor of importance to the job seeker, hence we assume these individuals would accept a job based on salary, cost of living, and disposable income. Local crime levels and natural hazard risks are not of importance to them. A counterfactual case is a job seeker, who recognizes and considers whether their economic welfare could be affected by hazard risks, which are negative amenities associated with a job offer. These individuals include crime and natural hazards in their assessment by weighing the impacts of different types of hazards (personal (crime) vs. nature-based (natural hazards)) in their job search. In a simulation of sequential job elimination, the participant reduces the chance of selecting a job with unattractive job attributes, i.e., an embedded crime or natural hazard risk is associated with a specific offer. In particular, the counterfactual assumes there is a negative correlation between job choice and overall hazard risk.

Previous analyses have represented job choice in a discrete choice RUM in making choices to attain the greatest utility from the choices available to them (Louviere 2001, McFadden 1974). The RUM framework is well suited for modeling job attributes, job selection, and location decisions. A related approach that accounts for the heterogeneity of individuals is a stochastic model to represent a probabilistic estimate of the duration of unemployment (Kiefer 1988, Lancaster 1997). These models focus on the timing of state transitions and treat exit from unemployment as the outcome of an arrival process of job offers and acceptance decisions. RUM provides a clear microeconomic interpretation of preferences over job attributes, while duration models capture dynamic aspects of search, such as waiting, learning, and state dependence. Unlike the other duration models, in our model, we use the RUM as the basis for a multiperiod semiparametric Cox hazard model that as time evolves, less desirable offers are eliminated in the process. Job search depends on a collection of simulated job offers in a prospective analysis of alternative job tasks and location attributes and amenities.

There are applications of spatial economic models that include crime and natural hazard risks and their impacts on labor and housing markets (Fajgelbaum and Gaubert 2020). These microeconomic and macroeconomic models treat the hazards risks as a disamenity that can affect estimates of the benefits of infrastructure development (Balboni 2025), geographic barriers to mobility (Kennan and Walker 2011) and socioeconomic impacts of migration (Carstensen et al 2020). There also is empirical evidence that suggests that wealthier cohorts at the country scale are more likely to invest in self-protection, self-insurance, and or market insurance (Kellenberg and Mobarak 2008, Viscousi and Zeckhouser 2006). This economic behavior exhibits an increasing risk aversion to natural hazards as a motivation to protect one's own safety and wealth. These models assume that disamenities such as crime and natural hazards are idiosyncratic risks in uncovering disparities in amenity quality across individuals (Kennan and Walker 2011).

Economics research related to natural hazards has focused on the personal, community, and macroeconomic impacts and losses from potential disasters. Losses can be reduced through self-protection, self-insurance, market insurance, and public investment (Kousky et al 2006, Shavell 2014, Viscousi 1992, Lewis and Nickerson 1989, and Ehrlich and Becker 1972). Analyses have been undertaken that identify whether individuals would invest in risk reduction measures from natural hazards and crime that, given available hazard information, are not totally unexpected and unpreventable (Vicousi and Zeckhauser 2006). There have been a variety of studies comparing people who choose to live in areas at greater risk of hurricane loss, such as along the coast, with those who live inland (H. John Heinz III Center for Science, Economics and the Environment 2000). In addition, large cities attract more affluent consumers that, along with amenities, increase the salary demand and influence job choice (Howard and Liebersohn 2025). Individuals move across space because of differences in wages in small and large cities that cause net migration to high-wage locations. Surveys have shown that individuals move for jobs or because of housing costs (Molloy et al 2011).

On the other hand, natural hazards reduce market values and have negative financial impacts over the long term (Ratcliffe et al 2019). A national survey concluded that four-fifths of the respondents favored government assistance for victims of natural disasters, but only one-third favored government assistance for victims when natural disasters happened to people living in high-risk areas (Vicousi and Zeckhauser 2006). In the national survey, policy preferences for disaster relief reflect both compassion for the unfortunate, and for self-interest.

Economic models have been used to communicate the benefits and costs of hazard insurance to minimize individuals' disaster losses (Husted and Nickerson 2019, Grossi and Kunreuther 2005, Kunreuther and Pauly 2004). Natural and crime hazards disamenities have been included in hedonic models affected by earthquake land use regulation (Brookshire et al 1985, Bishop 2012, Bayer et al 2009); and benefit – cost analyses for regulation and mitigation investment using an earthquake damage scenario simulation under different mitigation policies (Bernknopf and Amos 2014). Short run hazard communication of natural hazard information has been demonstrated as a positive economic impact on high-risk locations (Economou et al 2016, Restrepo 2017). Based on these analyses, it is evident that crime levels and natural hazard risks are disamenities that affect a $FJC$.

## 3. Economic Model and Hazard Modeling Framework

The foundation of our approach is a RUM with time varying preferences that we use to evaluate the latent utility associated with hazard risks. Jobs with lower utility are more likely to be eliminated. Elimination of job offers occurs sequentially across rounds, and the last surviving job becomes the final job choice. The RUM includes random disamenities that generate probabilistic elimination decisions that over time create an event history. The history defines a series of decisions in a discrete-time Cox hazard modeling framework. This model can be estimated as a discrete choice logistic regression (Allison, 1982). The Cox model is an estimate of the probability of elimination conditional on the remaining set of jobs in a choice set. Following Allison (1982), we link together a series of discrete choices based on the RUM logit probabilities with a flexible statistical framework of competing risks among job alternatives (Han and Hausman, 1990). Together, these results provide a direct link between survival analysis and discrete choice models. We build upon this relationship. Instead of using a Cox Proportional Hazard model that assumes a constant hazard rate over time (Allison 1982), we estimate a semiparametric version of the Cox model in an ECH model (Han and Hausman 1990). The flexibility of the ECH model accommodates the need for a complex natural hazards covariate that is both time dependent (latent time related preferences) and time independent (geographic exposure to natural hazards risks). We demonstrate that the natural hazards effect has a significant effect on a final job choice, albeit later in the search process. Based on the results of the seek-and-screen process, we develop an application about the elimination of specific jobs in a sequential hazard representation of discrete choice.

The model is based on three premises that underlie a job search decision: The labor demander and supplier have compatible incentives, individuals are aware and expect a government safety net, and an individual's risk perception about natural hazards and crime are disamenities. First, it is assumed that there is an incentive compatibility between an employer and an employee regarding disaster preparedness and response that motivates ensuring mutual safety, operational continuity, and economic security. Individuals perceive their fiscal responsibility and the role of public institutions could be affected by natural hazards (Balboni 2025, Lewis and Nickerson 1989). Second, there is an assumption that FEMA and other government agency policies will provide public assistance to limit the individual financial burden of self-protection and self-insurance (Kousky et al 2006, Lewis and Nickerson 1989, Yezer 2010). Third, there are stated and revealed preferences for the willingness to pay for risk reduction (Landry et al 2011, Viscousi and Zeckhauser 2006, Bernknopf et al 2003, Brookshire et al 1985). In these studies individuals act as we would anticipate, they expect that a natural or environmental hazard lowers property values in hazard zones and that higher incomes compensate for the willingness to accept higher risk locations and increased employment hazard exposures.

The adaptive discrete choice model has four stages: job seeker evaluation to identify preferences, sequential elimination, estimation of time varying preferences, and making a final job choice.

### A. Stage 1: Job seeker evaluation

The RUM job elimination decision derives from a latent utility comparison among job offers (Louviere 2001). Each job $j$ has latent utility that can be represented by a discrete choice utility model:

$$(1) \quad U_{ij} = V_{ij} + \varepsilon_{ij} = \beta_1 (Sal/CoL)_j + \beta_2 CR_j + \beta_3 NHR_j + \varepsilon_{ij}$$

where $U_{ij}$ is the total utility individual $i, i = 1, \dots, I$, gets from job $j = 0, \dots, J$, $V_{ij}$ is the systematic part of job utility, $\varepsilon_{ij}$ is the random component that are the unobserved preferences for choice $j$ held by individual $i$. $(Sal/CoL)_j$ where $Sal$ is salary associated with $j$, $CoL$ is a cost-of-living index where $j$ is offered. $CoL$ includes housing costs, food, transportation, health care costs, other necessities, and taxes (See https://www.epi.org/resources/budget/). $CR_j$ is the percentile of the crime rate (i.e., number of crimes per 1000 persons in a given city) relative to all crime rates in the US and is assigned to a category from 1 (Low – Bottom 10%) to 4 (High – Top 10%) as assigned from National Incident Based Reporting System (FBI 2022). $NHR_j$ is the risk potential for negative impacts from natural hazards by county for $j$. A hazard risk value is assigned from the FEMA National Risk Index (NRI) and takes into consideration the individual hazard risks, where $\sum_{NRI=0}^{6} NHR_{NRI(j)}$, of $j$ (FEMA 2025). These job attributes determine latent utility.

$V_{ij}$ motivates both the RUM utility choice probabilities and the Cox hazard model in stage 2. In the model, it is the observable component of utility that converts job attributes into a single index determining both choice probabilities and elimination risk. Elimination risk is estimated in a Cox hazard model, so that

$$(2) \quad V_{ij} = X_j\beta = \beta_1(Sal/CoL)_j + \beta_2 CR_j + \beta_3 NHR_j$$

**B. Stage 2: Sequential elimination**

The RUM is implemented as a process of sequential elimination, where the probability of elimination is the hazard associated with a binary choice. The Cox hazard model is estimated, which is equivalent to a series of conditional logit probabilities of job removal from a choice set obtained by way of simulation. In the Cox hazard model, each $j$ has an instantaneous hazard rate $h_j$:

$$(3) \quad h_j = \exp(X_j\beta)$$

The probability that job $j$ is chosen by individual $i$ to be eliminated next from choice set $C$ is

$$(4) \quad P_{ij} = \frac{h_j}{\sum_{n\in C}(h_n)}$$

So, to make relative comparisons between offers from the choice set is to estimate the log odds:

$$(5) \quad \log\frac{P_{ij}}{P_n} = V_{ij} - V_n$$

Probabilities depend only on differences in utility between $j$ and the remaining jobs in $C$. At each round $\mathrm{R}(n)$, the individual evaluates job $j$ and makes a binary decision (Allison 1982):

$$(6) \quad \gamma_{j\mathrm{R}(n)} = \begin{cases} 1 & \text{job is eliminated at } \mathrm{R}(n) \\ 0 & \text{job survives to next } \mathrm{R}(n) \end{cases}$$

where $\gamma_{j\mathrm{R}(n)}$ is the event indicator used to determine whether a job is eliminated or continues in the choice set. A job is eliminated if utility falls below a screening threshold:

$$(7) \quad \gamma_{j\mathrm{R}(n)} = 1, \text{ if } U_{ij} < \tau_{\mathrm{R}(n)}$$

where $\tau_{\mathrm{R}(n)}$ is a round specific screening threshold. Jobs are removed with a logit probability represented by the hazard rate:

$$(8) \quad h_j\big(\mathrm{R}(n)\big) = h_0\big(\mathrm{R}(n)\big)\exp\big(X_j\beta\big)$$

where $h_j\big(\mathrm{R}(n)\big)$ is the instantaneous hazard rate of elimination as a binary choice: eliminate = 1, or survive to the next round = 0 in a specific round, and $h_0\big(\mathrm{R}(n)\big)$ is the baseline for each job profile ID offer still available in $C$.

The Cox hazard model is an estimate of whether the binary elimination decision equals 1 at each $\mathrm{R}(n)$ and a job is eliminated:

$$(9) \quad \big(h_j\big(\mathrm{R}(n)\big)\big|X, \beta, h_0\big) = P\left(\gamma_{j\mathrm{R}(n)} = 1\middle| \text{job survived to } \mathrm{R}(n)\right)$$

where $\big(h_j\big(\mathrm{R}(n)\big)\big| \cdot\big)$ is the hazard of elimination in $\mathrm{R}(n)$. The series of choices creates the event history.

**C. Stage 3: Time varying preferences**

Time varying preference enters our model as the product, $NHR_j \times \mathrm{R}(n)$, which represents a change in an individual's risk assessment during the search process. In the approach, early rounds are interpreted as loose screening and later rounds are a stricter comparison among the top options. Hence, jobs with lower utility are more likely to be eliminated earlier, so they have a higher hazard rate.

The ECH model in equation 10 is used to evaluate the $NHR_j$ impact in the RUM. For comparison, we can assume a PH model in which we remove the interactive covariate $\left(NHR_j \times \mathrm{R}(n)\right)$ in equation 10:

$$(10)\ \ h_j(R(n)|\boldsymbol{X}, \beta, h_0) = h_0\big(\mathrm{R}(n)\big) \cdot \exp\Big(\beta_1 (Sal/CoL)_j + \beta_2 CR_j + \beta_3 NHR_j + \delta_1 \left(NHR_j \times \mathrm{R}(n)\right)\Big)$$

where $h_j(R(n)|\boldsymbol{X}, \beta, h_0)$ is the instantaneous hazard rate in $R(n)$ given salary and hazards covariates, $h_0\big(\mathrm{R}(n)\big)$ is the baseline hazard or reference case that includes salary and crime risk covariates (round-dependent), and $\exp\Big(\beta_1 (Sal/CoL)_j + \beta_2 CR_j + \beta_3 NHR_j + \delta_1 \left(NHR_j \times \mathrm{R}(n)\right)\Big)$ is the salary and hazards covariates (round-independent and round-dependent).

Elimination is estimated as the instantaneous value of a binary choice of 1 = accept, and 0 = reject, that determines the hazard rate for a job offer at $R(n)$. Both the PH and ECH are estimated. Table 4 below contains the results of the models. Then we compare them by testing whether there is a requirement for time dependency in the hazard model. If the interactive term $\left(NHR_j \times \mathrm{R}(n)\right)$ is insignificant, the requirement for application of the assumption of proportional hazards is satisfied (Kleinbaum 1996), and the PH model is appropriate for analysis. If the interactive term is significant, the proportional hazard assumption is not satisfied, and the ECH model is more appropriate. In the ECH model, $\beta$ coefficients are round independent and $\delta$ coefficients are round dependent.

**D. Stage 4: Making a final job choice**

The process is to seek, screen, eliminate and select a final job. A simulation is representative of a multi-round elimination process to find the job with the greatest realized utility. It is the job that survives the longest to become the $FJC$.

Measurement of the effect of negative amenities on job search in the Cox hazard models is contained in the hazard ratio ($HR$) of accepting or rejecting a job offer, during a specific time interval R($n$), where $n = 0, \dots, n$. If $HR = 1$, there is no effect, while if $HR < 1$, jobs that are identified at a specific hazard level survive longer, and if $HR > 1$, jobs are eliminated earlier in the search. A final job is affected by salary and cost of living and crime risk (reference case) and by salary and cost of living, crime risk, and location-specific natural hazard risks (counterfactual case).

Since our interest is a focus on whether an individual is concerned about the overall natural hazard risk in their $FJC$, the hazard ratios ($\widehat{HR}$) that are derived from the $\beta$ and $\delta$ coefficients in the PH and ECH models creates a $NHR$ effect on $FJC$ by round. $\widehat{HR}$ provides an estimate of whether subsequent job offers to the individual are unaffected at any given round has a greater, equal, or lower hazard rate of job elimination during the next round relative to an individual's choice from the baseline. Thus, $\widehat{HR}$ is a relative measure between job profile IDs and $h_0$ for determining the possible elimination of a job offer with less preferred amenities. $\widehat{HR}$ is measured as a time-dependent indicator (Kleinbaum 1996):

- Hazard Ratio = 1: An $\widehat{HR}$ equals one when the numerator and denominator are equal. This means there is equivalence between the reference and counterfactual cases, when both groups experience the same number of events in a round.
- Hazard Ratio > 1: An $\widehat{HR}$ is greater than one when the numerator is greater than the denominator in the hazard ratio. In this outcome the job profile experiences a higher failure probability within any given period than all job profile IDs.
- Hazard Ratio < 1: An $\widehat{HR}$ is less than one when the numerator is less than the denominator in the HR. Consequently, the job profile group experiences a lower failure probability during a round than all job profile IDs. $\widehat{HR}$ in equation 11a is for a PH and equation 11b is for a ECH model respectively:

$$\widehat{HR} = \exp(\beta_1(SaL/CoL) + \beta_2 CR + \beta_3 NHR) \tag{11a}$$

$$\widehat{HR}(\mathrm{R}(n)) = \exp\left(\beta_1(SaL/CoL) + \beta_2 CR + \beta_3 NHR + \delta_1\left(NHR_j \times \mathrm{R}(n)\right)\right) \tag{11b}$$

By using the coefficients in the Hazard Ratio, we can estimate the $NHR$ effect at a specific time that is conditioned on the prior simulation round.

The last input to the framework is the implementation cost for protection and preparedness ($PrCost$). $PrCost$ is based on the following question: How much would you be willing to spend to avoid harm or loss from the following hazards: earthquakes, wildfires, floods, landslides, severe weather, and volcanic activity? We consider the dollar values a rough approximation of an individual's willingness to purchase natural hazard loss mitigation. Dollar estimates for an individual's cost are put into five possible expenditure bins (i.e., \$0 - \$1,000, \$1,001 - \$10,000, \$10,001 - \$100,000, \$100,001 - \$1,000,000, \$1,000,000+) for the six types of natural hazards included in the choice experiment[2]. With the estimated cost of prevention and preparedness for overall natural hazard risk, we can forecast an individual's expected affordability for overall natural hazard risk associated with their $FJC$. We assume that the cost of the $NHR$ protection input is cost that can be chosen by the decision maker.

As there is no good single dataset that exists that details actual $PrCost$ for hazard loss at the county level, we created a financial burden indicator. The burden ratio represents an individual's cost of prevention and preparedness. The $PrCost$ is selected from the five expenditure bins provided above in the choice experiment question. We assume that an individual is more likely to choose a minimum prevention cost because over 75% of the participants' selections in Table 8 below would not spend more than \$10,000 to protect themselves. By combining an individual participant's expected disposable income and a selected estimated prevention cost, the consumer can identify an amount that could be spent between a minimum and maximum cost of prevention. In equation 12, we express the intent of an individual who chooses a minimum prevention cost for $BR$

$$(12)\ \ BR = \frac{disposable\ income - \text{minimum}\ prevention\ cost}{disposable\ income} \gtreqless 0$$

[2] We did not tie the question in the experiment to any reference of their current occupation nor to the job they chose, so we did not estimate an explicit willingness to pay for hazard mitigation. The question in the experiment gave no reference point for "this is how much you have to spend to protect yourself", prior to asking how much a participant would be willing to spend.

The $BR$ yields the proportion of an individual's income left over after paying the minimum prevention cost. The burden level is based on three $BR$ categories:

- Low – burden ratio ($BR > 0.5$) – an individual retains 50% or more of their disposable income,
- Moderate – burden ratio ($0 < BR < 0.5$) – an individual retains some but less than 50% of disposable income,
- High – burden ratio ($BR \leq 0$) – an individual either breaks even or has a negative disposable income.

For example, using equation 12, if an individual had a disposable income of $26,476, and the prevention cost they would expect to pay is between $10,000 – $100,000, the $BR$ for $10,000 is 0.62 and easily affordable (low burden), while the BR for $100,000 is -2.78, which would not be covered by disposable personal income (high burden).

## 4. Choice Experiment Materials

The aim of the choice experiment is to determine if participants use knowledge of natural hazards when making life choices, such as relocating to a new location. The set of job offers is based upon the participants' job search locations and their educational attainment and degree focus. These criteria are used to create the set of 16 job offers (i.e., job profile IDs) that meet the job list criteria in Table 1 from the full set of 38,411 job options. The dynamic of accepting a job involves how people weigh salary and cost of living data against risk factors, such as crime and natural hazards. Participants identify, assess, and consider these data as they screen attributes of specific job offers. In the experiment, participants progressed through the following major steps to reach a final analysis of their decisions: specifying job search parameters, choosing a job, reflecting on reasons for final job choice, assessing risk tolerance, and reflecting on reasons for choosing current location of residence and current job. Participants were allowed to participate in the experiment as many times as they wished.

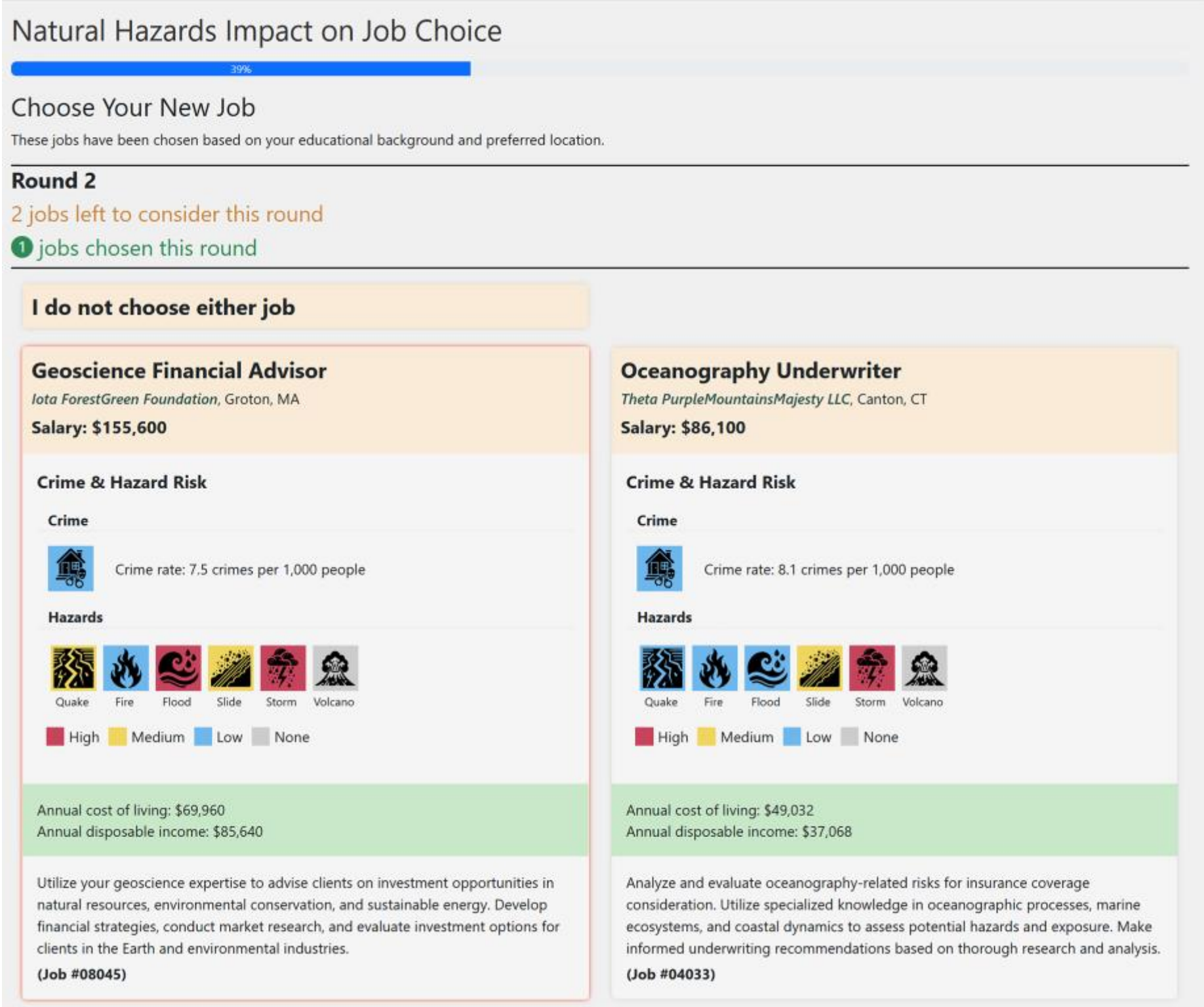

**Figure 1. Job Offer Details**

*Notes:* Job offer, job title, company name, city, state, salary, crime and natural hazard risk levels, crime risk level (color) and rate per 1000 people, natural hazard risk level for earthquakes, wildfires, floods, landslides, severe weather, and volcanoes, cost of living and disposable income, job description, and job profile ID number (Gonzales et al 2026).

The seek-and-screen process is initiated in the first round with 16 job offers randomly paired and presented to the participant pair by pair as shown in Figure 2 Column R(1). The participant can either choose one job offer or reject both job offers in the pair (see Figure 1). At the beginning of R(2) in Figure 2, the selected jobs from the previous round, R(1), are randomly paired and presented to the participant. If there is an odd number of offers selected in the previous round, then the first job offer in the round is also paired with the last job offer in the round. If a participant chooses a job from each pair of offers, then they will have selected 8 jobs in R(1), then select 4 jobs in R(2) and 2 jobs in R(3), in order to make a final choice by R(4). If a $FJC$ is not reached by the fourth round, the participant is presented with a new set of 16 job offers and the bracketed job choice selection process begins anew. A $FJC$ can be made at any time during the selection process. For example, a participant may choose one job from one pair in the first round and reject all the rest of the offer pairs in that round, thereby reaching the $FJC$ in the first round of selection.

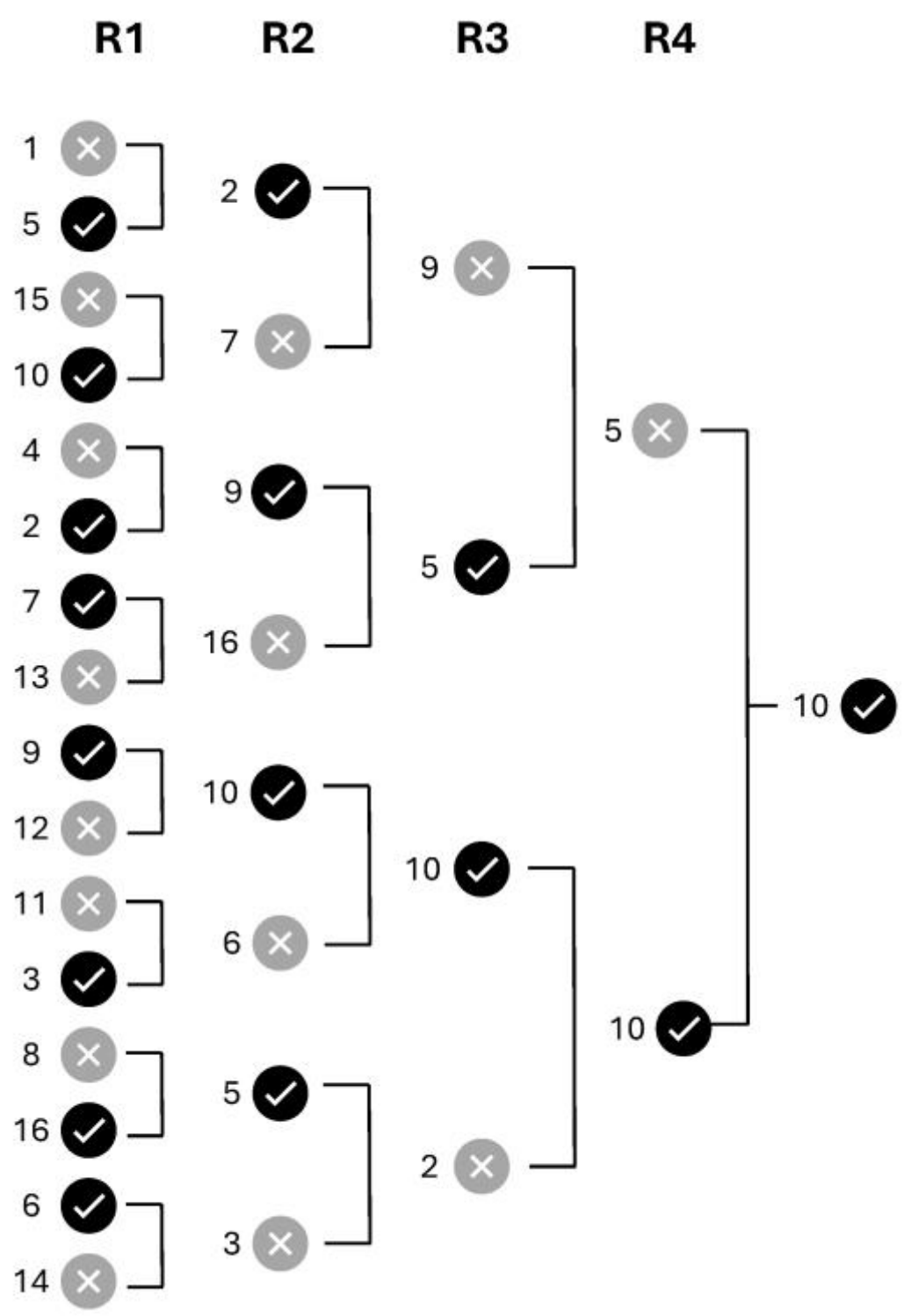

**Figure 2. Bracketed job choice selection process with randomization of selected jobs**
*Notes:* Example job choice brackets. "R" (n) indicates the round of job choice selection and is right censored. In each round, jobs selected in the previous round are randomly paired (Adapted from Gonzales et al 2026).

At the end of the experiment, the participant can provide reasons for their $FJC$, and whether disamenities influenced their decision-making process. The participant is asked about their level of concern with natural hazards in the location where they live, about how much they would spend to avoid loss and harm, and the severity of impacts they have experienced to gain insight into a participant's risk tolerance level. The participant also is asked about the factors that influenced their choice of current residence and the factors that influenced their choice of their current occupation. This information is used to understand the participant's preferences and priorities when choosing a job. Next, the participant is asked about their demographic information for calculating the individual's cohort. Finally, the participant is shown their final job choice and is provided with an analysis of their hazard risk tolerance and the importance of factors in choosing their current location of residence and $FJC$.

The seek-and-screen process is used to narrow the selection process to a final choice. Because each job offer provides information about salary, location, hazard risk level, crime risk, cost of living, and other attributes, participants can consider trade-offs between financial and personal safety, and other factors. Salary-to-cost of living, crime risk, and hazard risk are assigned a unique job profile ID, which ranged from 1 to 16 (Table 1). We varied these three variables to understand which combination would be most attractive to participants. Salary-to-cost of living ratio accounts for: 1) salary, 2) cost of living, and 3) disposable income all at once and is normalized across all jobs because it is a ratio. Stated preferences indicated that wages and job components strongly influence job choice, and that natural hazards have a small, yet statistically significant negative marginal impact on moves to higher cost and higher risk locations.

**Table 1—Job profile criteria**

| Job profile | Salary relative to cost | Crime risk | Natural Hazard |
|---|---|---|---|
| **1** | Med (1.21 - 1.5) | High (4) | High - Very |
| **2** | Low (1 - 1.2) | High (4) | Very Low (1) |
| **3** | High (1.51 - 5) | Low (1) | Med (3) |
| **4** | High (1.51 - 5) | Low (1) | High - Very |
| **5** | Low (1 - 1.2) | Low - Med (1 - 2) | Med - Very High |
| **6** | Med (1.21 - 1.5) | Low - Med (1 - 2) | Med - Very High |
| **7** | High (1.51 - 5) | Low - Med (1 - 2) | Med - Very High |
| **8** | Low (1 - 1.2) | Low - Med (1 - 2) | Very Low - Med |
| **9** | Med (1.21 - 1.5) | Low - Med (1 - 2) | Very Low - Med |

| 10 | High (1.51 - 5) | Low - Med (1 - 2) | Very Low - Med |
|---|---|---|---|
| **11** | Low (1 - 1.2) | Med - High (2 - | Med - Very High |
| **12** | Med (1.21 - 1.5) | Med - High (2 - | Med - Very High |
| **13** | High (1.51 - 5) | Med - High (2 - | Med - Very High |
| **14** | Low (1 - 1.2) | Med - High (2 - | Very Low - Med |
| **15** | Med (1.21 - 1.5) | Med - High (2 - | Very Low - Med |
| **16** | High (1.51 - 5) | Med - High (2 - | Very Low - Med |

In our choice experiment hazard analysis, we select participant records where a $FJC$ is chosen and censor participant records who do not make a final job choice. This yielded an analysis dataset of 535 participant records. On average, final job choices had higher salaries, lower crime risk, and higher hazard risk than their alternative option. Gonzales et al (2025) used a decision tree analysis on data from this experiment and showed that at a salary of $82,650 and below generally were unwilling to accept jobs with higher hazard risk; however, above this threshold, higher hazard risks were more acceptable. Our results in Table 2 show that as salaries increase, acceptance of higher hazard risk also increases, with the largest increase in hazard risk acceptance occurring when salaries exceed $100,000, suggesting there is a compensatory effect of salary.

**Table 2 — Differences in crime risk and hazard risk between final job choices and alternative rejected job offers**

| | Crime Risk Difference | | | Hazard Risk Difference | | |
|---|---|---|---|---|---|---|
| **Salary bin** | **decrease** | **no change** | **increase** | **decrease** | **no change** | **increase** |
| $30,000 to $40,000 | 0% | 100% | 0% | 100% | 0% | 0% |
| $40,000 to $50,000 | 11% | 43% | 46% | 64% | 21% | 14% |
| $50,000 to $60,000 | 9% | 50% | 41% | 41% | 41% | 19% |
| $60,000 to $70,000 | 18% | 53% | 29% | 36% | 42% | 21% |
| $70,000 to $80,000 | 38% | 38% | 25% | 21% | 49% | 30% |
| $80,000 to $90,000 | 54% | 35% | 11% | 11% | 55% | 34% |
| $90,000 to $100,000 | 45% | 42% | 12% | 14% | 64% | 23% |
| $100,000 to $100,000+ | 51% | 36% | 13% | 9% | 39% | 52% |

Econometric analysis of job search is populated with the outputs from the choice experiment. We use the results to estimate the systematic risks associated with local environmental disamenities. We incorporate the time variability in preferences toward environmental risks collocated with a job opportunity and how an individual has chosen to eliminate the hazard risks with each successive round in the simulation. The individual learns and updates their likes and dislikes as a latent effect of the specific negative amenities associated with job offers. This approach represents a time-dependent search as an adaptive risk assessment (Gati et al 2019, Gati and Asher 2001).

The modeling framework assumes that individuals make more informed choices as they move through rounds as they evaluate alternative job characteristics (Viscousi 1992). The dynamic construct of the experiment allows the participant to learn about how job attributes can affect their cost of living and choice of location relative to their employment aims. To set up the analysis, we graph the mean value for each of the factors at each round in Figure 3 to assess trends in acceptance and rejection of job offers across the rounds.

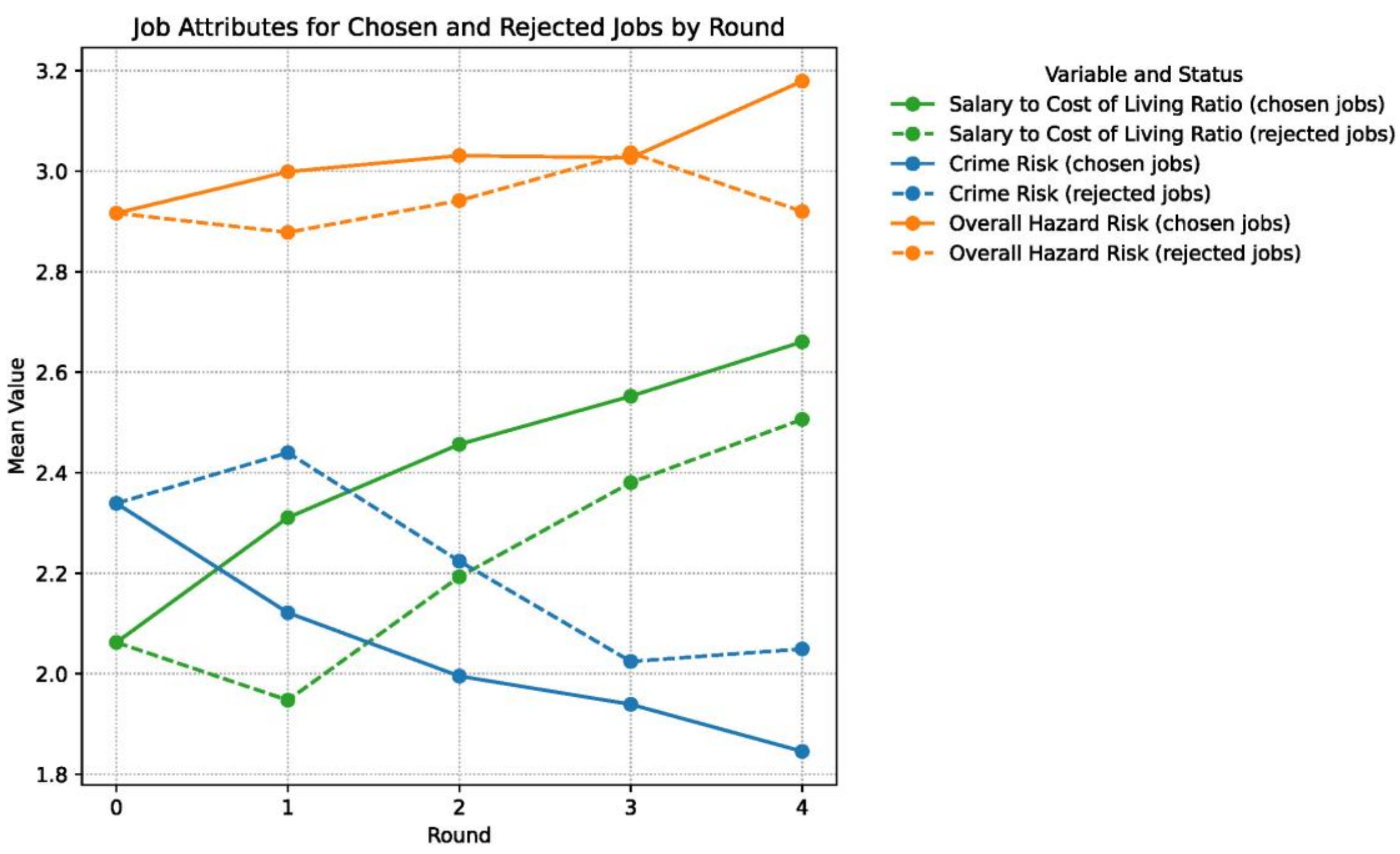

**Figure 3. Trend for job profile ID attributes for accepted and rejected jobs by round**

At the start of the selection process, we estimate the baseline hazard $h_0(\mathrm{R}(n))$ for each job profile ID with the $(Sal/Col)$, $(CR)$, and $(NHR) = 0$. The baseline hazard represents the reference case. The average $(Sal/Col)$ of chosen jobs at $h_4$ was 29% higher than the mean value at the start of the experiment, while the crime risk was 21% lower and the hazard risk was 9% higher. Translating the mean values into monetary units, salaries increased across decision rounds for all job offers from $74,162 to $98,327, for job profiles that were chosen in any round the average salary increased from $82,284 to $110,436, and for the $FJC$ selections, the average salary was $100,885. This outcome suggests a consistent preference for higher-paying jobs across rounds. When examining the survival rate of jobs (Table 5), jobs with high salary relative to cost of living regardless of crime risk and hazard risk were consistently retained at a higher rate than the baseline survival rate for all jobs. However, high salary jobs with lower crime risk had the highest survival rates of all jobs, while those high salary jobs with higher crime rates, though outperforming the baseline survival rate, had lower survival rates than high salary jobs with lower crime rates. Low salary jobs had a 5% to 11% survival rate, with higher survival rates associated with lower hazard risk, but not lower crime risk. The largest changes for crime risk occurred between R(1) and R(3), since accepted jobs were more likely to be chosen in lower-crime areas. That is, on a scale from 1 (low) – 4 (high), the accepted job average crime level was 1.98, while rejected jobs had an average value of 2.20. Hazard risk remained near 3.0 for rounds 1 through 3, but increased in the final round by 0.15, signaling a slope reversal as a function of location and time dependent preference for $NHR$ chosen and rejected jobs. This increase in hazard risk between R(3) and R(4) indicates that an increased $NHR$ is acceptable when making a final job choice. Analysis of the larger set of data from the GRANDE project indicates that even among those who said hazard risk was influential in their final job choice, people often chose jobs with moderate- to high hazard risk, with severe weather being the most accepted hazard of all hazard types (Gonzales et al 2025).This slope sign change in natural hazard risk between R(3) and R(4) is a clear signal to test whether the assumption of constant proportionality in a PH model is appropriate or is it better to use an ECH model.

**Table 3 — Mean salaries for all job offers and accepted job offers by round**

| | Mean salaries for jobs | |
|---|---|---|
| **Round** | **all job offers** | **accepted job offers** |
| **0** | $74,162 | $82,284 |
| **1** | $82,448 | $90,344 |
| **2** | $89,791 | $98,478 |
| **3** | $98,327 | $110,436 |

## 5. Analysis and Results

Our analysis includes estimation of Kaplan-Meier (KM) conditional probabilities of survival, and a Cox hazard model. The objective is for an individual to accept a job offer and eventually make a final job choice. In their search, the individual identifies the job profile IDs to be eliminated $(h_j)$ at a specific time, or if no choice is made, they are censored at the end of the simulation. For analysis of the hazard models, the baseline is a PH model. In this case $h_j(\mathrm{R}(n))$ is based on time independent covariates that are average values for salary, crime level and natural hazards through all the rounds in a simulation. In the ECH model, $h_j(\mathrm{R}(n))$ depends on time independent and time dependent covariates. Both models are assumed to have a positive correlation for salary, and a negative correlation between job choice crime levels and overall natural hazard risk. But unlike the crime level, natural hazards risks also are time-dependent, albeit late in the process.

The way the model is constructed allows for learning about specific job characteristics in consecutive attempts to simulate the dynamics of the tradeoffs among salary, cost of living, crime, and natural hazards. Job location occurs as existing, potential, and $FJC$ locations. If a final job choice is not found by $\mathrm{R}(4)$, the individual is censored for choices for a particular $\mathrm{R}(1) - \mathrm{R}(4)$ period. The individual is not excluded per se, their specific set of choices are. Following the individual's censored choices, they can start the search over with a fresh set of choices.

The KM probability estimate is a product of the conditional probabilities observed for all 4 rounds[3]. Each term in the product is the probability of exceeding a specific ordered failure time $\mathrm{R}(n)$ given that a job profile ID survives to that failure time and is the cumulative effect of the hazard as they progress through the rounds (Kleinbaum 1996). One outcome of the statistical analysis yields the KM conditional probability that an $FJC$ is not eliminated over the 4 rounds. In Figure 4 we illustrate the KM survival rate for the 16 job profile IDs through R(4).

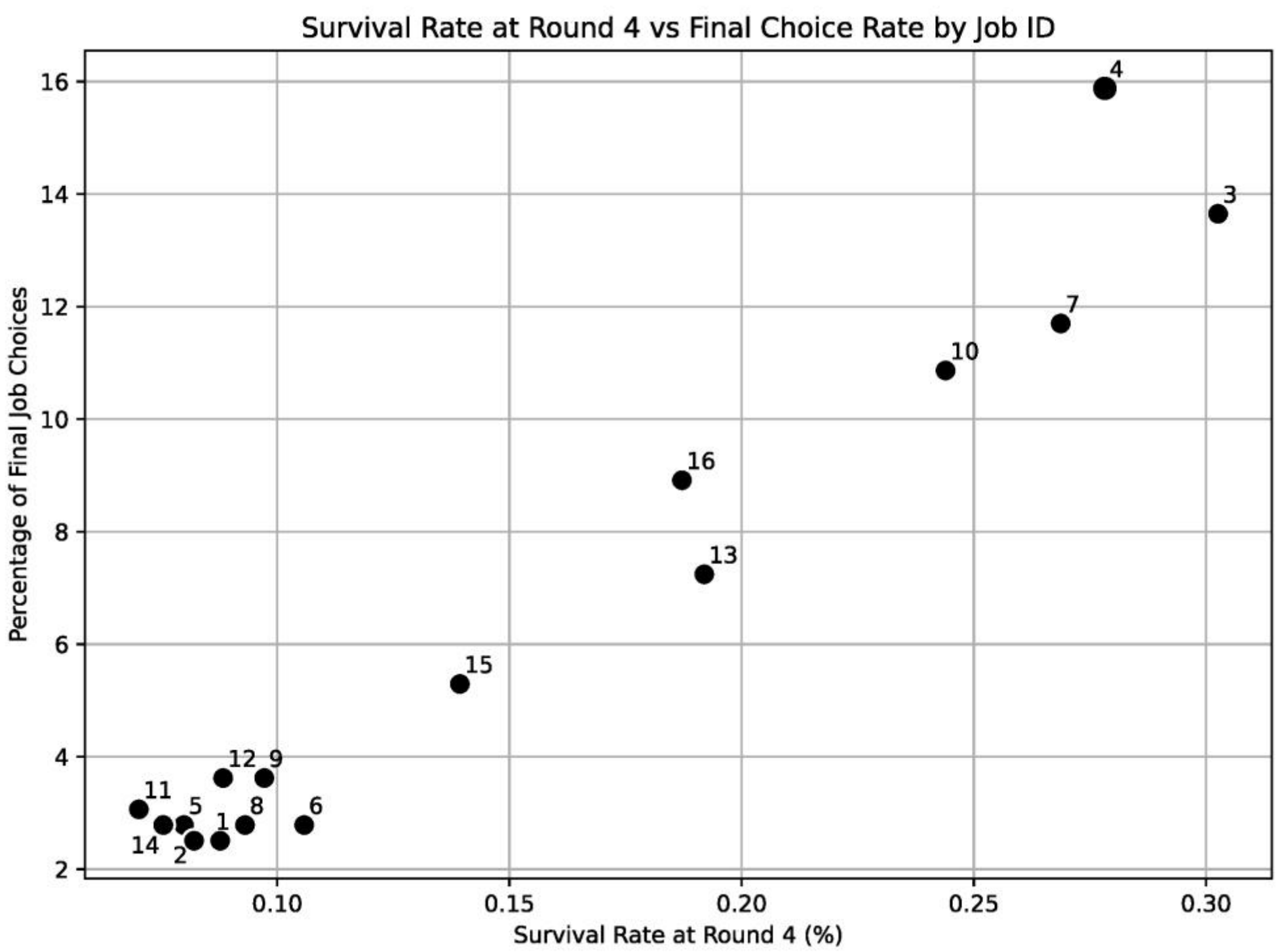

**Figure 4. Kaplan-Meir survival rate for jobs compared to the FJC by job profile ID by percentage of final job choices per job profile ID**

*Notes:* However, KM probabilities tell us little about the specific effect that NHR has on the job elimination of an offer.

Figure 4 depicts the interaction of the survival rate and the percentage of $FJC$s for each job profile ID. Job profile IDs clustered near the origin are eliminated early in the process because they are lower paying and have higher hazard risk, which could be due to location factors. While, job profile ID 3 had a higher survival rate, job profile ID 4 was more frequently chosen as the $FJC$.

[3] The probability is calculated by multiplying the conditional probabilities of survival at each round considering participants job offers at that time. The result is the conditional probability of surviving at $\mathrm{R}(n)$ by dividing the number of participants still seeking the best job offer at $n$ by the number of participants still looking for a job just before $n$.

To evaluate the $NHR$ effect, we estimate the ECH model in equation 10 to evaluate the impact on the RUM. In Table 4 we provide the econometric analysis for both the PH and ECH models. Analysis shows that the salary to cost of living ratio ($Sal/CoL$) and crime rate ($CR$) are significant predictors of job elimination, whereas in the PH model, natural hazard risk ($NHR$) is not significant when constrained to have a constant effect through the rounds ($NHR{:}\, \beta = -0.01, p = 060$).

Because inspection of Figure 3 suggested that the proportional hazards assumption may be violated for $NHR$, we estimated the ECH model by adding the time-dependent interaction term $\left(NHR \times \mathrm{R}(n)\right)$, where $\mathrm{R}(n)$ is the round number. In the ECH model, $Sal/CoL$ ($\beta = -0.41, p < 0.005$) and $CR$ ($\beta = 0.12, p < 0.005$) remained significant and are similar in magnitude to the PH model, while $NHR$ became significant ($\beta = -0.23, p < 0.005$) and the interaction term also is significant and positive ($\delta = 0.07, p < 0.005$). This indicates that the effect of NHR changes across rounds rather than remaining constant. The estimated total $NHR$ effect is given by $\beta_{NHR}\left(\mathrm{R}(n)\right) = -0.23 + 0.07\,\mathrm{R}(n)$, which implies the $NHR$ effect becomes less negative over time. Model fit statistics also favored the ECH model over the PH model (a trial log-likelihood: -53710.21 vs. -53882.88; partial AIC: 107488.42 vs. 107771.77; concordance: 0.63 vs. 0.61), and the increase in the log-likelihood ratio test statistic (694.05 on 4 degrees of freedom (df) vs. 408.7 on 3 degrees of freedom) supports the conclusion that the PH assumption is not satisfied for $NHR$ and that the ECH model provides a more appropriate specification for geospatial job-choice data.

**Table 4 — Cox Proportional Hazard and Extended Cox Hazard models result for the Natural Hazards and Job Choice seek and screen discrete choice simulation.**

| | |
|---|---|
| Number of observations | 8586 |
| Number of events observed | 6452 |

| | | Variables | | | |
|---|---|---|---|---|---|
| **Model** | **Estimate** | **($S/CoL$)** | ***CR*** | ***NHR*** | **$NHR \times g(\mathbf{R})$** |
| **Cox Proportional Hazard (PH)** | $\beta$ | -0.4 | 0.13 | -0.01 | |
| | se ($\beta$) | 0.04 | 0.01 | 0.01 | |
| | Z | -13.34 | 8.73 | -0.53 | |
| | P | <0.005 | <0.005 | 0.60 | |
| | $(\exp^{\beta})^2$ | 0.67 | 1.13 | 0.99 | |
| **Extended Cox Hazard (ECH)** | $\beta, \delta$ | -0.41 | 0.12 | -0.23 | 0.07 |
| | se ($\beta$), se ($\delta$) | 0.03 | 0.01 | 0.02 | 0.004 |
| | Z | -13.50 | 8.88 | -12.39 | 16.17 |
| | P | <0.005 | <0.005 | <0.005 | <0.005 |
| | $(\exp^{\beta})^2$, $(\exp^{\delta})^2$ | 0.67 | 1.13 | 0.79 | 1.07 |

| Model | PH | ECH |
|---|---|---|
| partial log-likelihood | -53882.88 | -53710.21 |
| Concordance | 0.61 | 0.63 |
| Partial AIC | 107771.77 | 107488.42 |
| log-likelihood ratio test | 408.70 on 3 df | 694.05 on 4 df |
| -log2(p) of ll-ratio test | 290.8 | 492.21 |

The $\beta$ and $\delta$ coefficients in Table 4 are used to calculate $\widehat{HR}(\mathrm{R}(n))$ in equations 11a and 11b that are listed in the last two columns of Table 5. Participants considered natural hazard information to have a negative impact on $FJC$. The negative sign in the ECH on the $NHR$ confirms this relationship. However, the positive sign on the round dependent $NHR$ variable indicates that because the estimated coefficient $\hat{\delta} > 0$, $\widehat{HR}(\mathrm{R}(n))$ will increase exponentially with time (Kleinbaum 1996). Setting the $NHR$ effect to zero, we expect the sign for the effect to flip to positive between R(3) and R(4) at $\mathrm{R}(n) \approx 3.6$. Moreover, the sign flip to positive means there is a structural time varying effect that is not proportional and increases with time, for jobs under serious consideration for a $FJC$. Thus, the combined $NHR$ effect suggests that participants would consider natural hazards a manageable risk in places with higher salaries and less crime.

Table 5 contains KM survival probabilities and the PH and ECH Hazard Ratios for the 16 job profile IDs listed in Table 1. The KM values indicate the probability of survival up to $\mathrm{R}(n)$ relative to a baseline of all jobs. KM probabilities are a direct measure of the perceived risk that a job offer would survive and result in a final job choice by the fourth round. The first interval begins at R(0) and ends just prior to R(1). The survival probabilities for the remaining rounds are listed in columns 2,3, and 4 in Table 5. For example, the cumulative probability indicates that the baseline hazard has a probability of 0.157 of surviving until R(4). We compare the survival rate of each job profile ID (1 to 16) with that of the baseline in Table 1. Job profile ID's preferred to the baseline that have higher R(4) survival probabilities in Table 1 are: 3 = 0.308, 4 = 0.268, 7 = 0.263, 10 = 0.242, 13 = 0.199, and 16 = 0.185.

We compare the values for each model result and their hazard ratios in Table 5. Based on the analysis, all of the job profile IDs are fairly consistent across models. Job choices using the ECH model, which is the most robust model, identifies the top 6 job profile IDs to be 3, 4, 7, 10, 13, and 16, which are highlighted in bold. Relative to each job profile ID's baseline, all of the top 6 job profile IDs hazard ratios increased. However, none of them increased to greater than $HR = 1$, that would have suggested elimination earlier. Only job profile ID 2 had both $h_0$ and $HR > 1$, which means earlier elimination. As shown in Table 6, sequential elimination revealed that $FJC$s, are distributed across all rounds. The table shows round by round the percentage of participants' final job picks. By the end of Round 3, 56% of participants had made a $FJC$, and by the end of Round 4, the remaining 44% made a $FJC$.

**Table 5 — Kaplan-Meier survival probabilities, Hazard Ratio for the PH model $\widehat{HR}$, Hazard Ratio for the ECH model $\widehat{HR}(\mathrm{R}(n))$**

| | Kaplan-Meier Survival Probability (KM)* | | | | | | PH | | ECH | | Instant Hazard Rate |
|---|---|---|---|---|---|---|---|---|---|---|---|
| **Job ID** | **R2** | **R3** | **R4** | **Cond. $p(R2\|R1)$** | **Cond. $p(R3\|R2)$** | **Cond. $p(R4\|R3)$** | $h_0$ | $\widehat{HR}$ | $h_0$ | $\widehat{HR}(\mathrm{R}(n))$ | $h_0$* $\widehat{HR}(\mathrm{R}(n))$ |
| **All** | **0.43** | **0.24** | **0.16** | **0.43** | **0.55** | **0.65** | **0.71** | **0.71** | **0.71** | **0.71** | **0.53** |
| 1 | 0.37 | 0.17 | 0.09 | 0.37 | 0.47 | 0.50 | 0.94 | 0.91 | 0.93 | 0.93 | 0.87 |
| 2 | 0.27 | 0.12 | 0.10 | 0.27 | 0.44 | 0.80 | 1.04 | 1.03 | 1.04 | 1.02 | 1.06 |
| **3** | **0.59** | **0.43** | **0.31** | **0.59** | **0.73** | **0.72** | **0.51** | **0.52** | **0.51** | **0.52** | **0.27** |
| **4** | **0.57** | **0.37** | **0.27** | **0.57** | **0.65** | **0.72** | **0.50** | **0.51** | **0.50** | **0.52** | **0.26** |
| 5 | 0.34 | 0.14 | 0.07 | 0.34 | 0.42 | 0.49 | 0.81 | 0.79 | 0.81 | 0.79 | 0.63 |
| 6 | 0.46 | 0.22 | 0.14 | 0.46 | 0.48 | 0.64 | 0.73 | 0.72 | 0.73 | 0.72 | 0.52 |
| **7** | **0.55** | **0.37** | **0.26** | **0.55** | **0.68** | **0.71** | **0.56** | **0.57** | **0.56** | **0.58** | **0.33** |
| 8 | 0.35 | 0.18 | 0.11 | 0.35 | 0.51 | 0.58 | 0.80 | 0.80 | 0.80 | 0.79 | 0.64 |
| 9 | 0.44 | 0.21 | 0.11 | 0.44 | 0.48 | 0.52 | 0.72 | 0.72 | 0.72 | 0.71 | 0.51 |
| **10** | **0.53** | **0.33** | **0.24** | **0.53** | **0.63** | **0.72** | **0.56** | **0.57** | **0.56** | **0.57** | **0.32** |
| 11 | 0.32 | 0.11 | 0.05 | 0.32 | 0.35 | 0.47 | 0.89 | 0.87 | 0.88 | 0.87 | 0.77 |
| 12 | 0.41 | 0.18 | 0.10 | 0.41 | 0.45 | 0.55 | 0.80 | 0.78 | 0.79 | 0.79 | 0.62 |
| **13** | **0.47** | **0.29** | **0.20** | **0.47** | **0.61** | **0.69** | **0.60** | **0.62** | **0.60** | **0.62** | **0.37** |
| 14 | 0.33 | 0.14 | 0.09 | 0.33 | 0.43 | 0.61 | 0.89 | 0.88 | 0.89 | 0.87 | 0.77 |
| 15 | 0.41 | 0.20 | 0.13 | 0.41 | 0.49 | 0.65 | 0.80 | 0.79 | 0.79 | 0.79 | 0.62 |
| **16** | **0.49** | **0.28** | **0.19** | **0.49** | **0.57** | **0.66** | **0.59** | **0.61** | **0.59** | **0.61** | **0.36** |

*Notes:* Job ID is the job profile ID. Job ID All represents all jobs. R2, R3, and R4 refer to Round 2, Round 3, and Round 4 respectively. Cond. $p$ is short for Conditional $p$. *Highest probability or surviving or least chance of job profile ID elimination until R($n$) = 4. Top jobs are shown in bold.

The finalist round of job choices for the top 6 job profile IDs (3, 4, 7, and 10, 13, and 16) are listed in Table 7. The table is an assessment of the strengths of the final job choices over the other jobs in the choice set in the final round. We compared the mean difference in salary to cost of living, crime risk, and hazard risk between the job chosen versus the alternative in the pair. All $FJC$s had higher salary to cost of living ratios than the alternatives, and for job profile IDs 3 and 4, the crime risk was one category lower on average than the alternative job offer, while job profile ID 13 increased by a half step, and job profile ID 16 by three fourths of a step, but neither were enough to change the category. In terms of hazard risk, job profile ID 4 had a 1 step increase in hazard risk relative to the alternative, and job profile ID 7 had a 0.73 step increase relative to the alternative job offer. Other job profile IDs with increased hazard risk relative to the alternative offer included job profile ID 13 and job profile ID 3. Job profile IDs with lower hazard risk relative to the alternative included job profile IDs 10 and 16.

**Table 6 — Participant Final Job Choices by Round in the Adaptive Choice Experiment**

| Round | Participants Reaching an $FJC$ | Percent of all Participants |
|---|---|---|
| **1** | 68 | 13% |
| **2** | 94 | 18% |
| **3** | 140 | 26% |
| **4** | 233 | 44% |

**Table 7 — Mean differences between final job choices and the alternative job offers rejections for the top job profiles in R(4)**

| Job profile ID | Δ Salary to | Δ Crime Risk | Δ Hazard Risk |
|---|---|---|---|
| **3** | 0.36 | -1.09 | 0.04 |
| **4** | 0.47 | -1.00 | 1.19 |
| **7** | 0.74 | -0.15 | 0.73 |
| **10** | 0.65 | -0.30 | -0.47 |
| **13** | 0.81 | 0.50 | 0.60 |
| **16** | 0.89 | 0.76 | -0.50 |

Table 7 shows an increase in the salary to cost of living ratio for all 6 job profile IDs, lower crime rates (except for job profile IDs 13 and 16 where it rises slightly), and increased natural hazard risk (except for job profile IDs 10 and 16). Of the top jobs, job profile IDs 3, 4, 7, 13 had medium to very high levels of hazard risk, while job profiles 10 and 16 had very low to medium hazard risk. Relative to other job profile IDs, job profile IDs 3 and 4 showed smaller gains in salary with larger drops in the crime risk, and mixed changes in hazard risk. On the other hand, job profile IDs 7, 10, 13, and 16 had larger salary increases, mixed changes in crime risk, and only job profile IDs 10 and 16 had half step negative changes in hazard risk. This trend suggests that favorable changes in salary and higher natural hazard risk disamenities are more prevalent in preferred locations. This result is due to the combined $NHR$ effect that suggests participants would consider natural hazards a manageable risk in places with higher salaries and less crime. However, in the case of job profile IDs 13 and 16, higher crime was also acceptable with higher salaries.

## 6. Protection, Financial Burden and Forecasting Affordability

The participant conducts the simulation based on their individual preferences and by the last round have eliminated or limited the impact of disamenities. After the final job is chosen, individuals in the choice experiment are asked to rank a set of factors that influenced their decision to choose that job. They are then asked to reflect on their current situation, including ranking factors important to their choice of current residence and current occupation, concern for natural hazards at their current location of residence, experience of natural hazard impacts, and their willingness to pay for protection from hazards (Gonzales et al 2026, Slovic et al 2004). However, up to this point, decisions have been made without regard to the cost of protection.

The prevention cost question in the choice experiment provides the participant with the ability to consider a minimum (maximum) investment to avoid loss and harm from natural hazards. While this is decoupled from the $FJC$ information in the experiment, meaning that the participant is not provided specific information about the $FJC$ salary, disposable income, and hazard risk while considering their response to this question, the decoupling allows us to examine how realistic the participant is in estimating their ability to pay for managing the hazard risk. In other words, is it worth the financial burden to take precautionary measures and still choose to move to a high income and high-risk location, and how manageable is the affordability of prevention? We approach this decision with the seek-and-screen model through a measure of financial burden. Note that for this part of the analysis, we used a subset of our analysis dataset where the prevent cost question was completed. This yielded data from 350 participant records.

We use our results to develop an indicator of the affordability of natural hazard prevention and preparedness. The indicator is an anticipatory metric that allows an individual to scan many job options and identify how much it will burden them financially to mitigate the overall natural hazard of a job offer. We assume that a risk prevention decision is made with existing knowledge of public agency past investment in hazard protection and disaster reduction and relief. Our analysis employs the expenditure for prevention and preparedness, the salary-to-cost of living ratio, and the burden ratio to estimate the affordability of investing in hazard protection.

First, we calculate $BR$ from equation 12. Cost of prevention and preparedness provides input for 5 expenditure bins covering a wide range of investment costs in loss prevention. They range from no investment to greater than $1,000,000. To help with classifying whether the investment in protection is affordable, we divide participants' job choices into 3 broad disposable income categories that are delineated as: low, medium, and high. Even though the participant does this estimation external to the choice experiment we want to provide a quick way to check on whether an individual potentially could spend

an amount of money for safety they think they will need. The second step is to estimate an Anticipated Affordability ratio ($AA$). Here we use the participants in the choice experiment as an example. Table 8 contains the distribution of participant investment choices for 5 expenditure levels for each of the 6 natural hazards in the job search that shows that some investment is likely across all natural hazards.

**Table 8 — Participant selections for hazard risk expenditures by hazard type**

| | | Participants' Cost of Prevention and Preparedness Selections by Natural Hazard | | | | | |
|---|---|---|---|---|---|---|---|
| **Expenditure Bin** | | **Earthquake** | **Wildfire** | **Flood** | **Severe** | **Landslide** | **Volcano** |
| $0 to $1,000 | n | 203 | 174 | 145 | 116 | 240 | 291 |
| | % | 58% | 50% | 41% | 33% | 69% | 83% |
| $1,000 to $10,000 | n | 93 | 103 | 124 | 139 | 68 | 30 |
| | % | 27% | 29% | 35% | 40% | 19% | 9% |
| $10,001 to $100,000 | n | 39 | 61 | 65 | 75 | 32 | 17 |
| | % | 11% | 17% | 19% | 21% | 9% | 5% |
| $100,001 to $1,000,000 | n | 7 | 4 | 9 | 9 | 4 | 6 |
| | % | 2% | 1% | 3% | 3% | 1% | 2% |
| $1,000,001+ | n | 8 | 8 | 7 | 11 | 6 | 6 |
| | % | 2% | 2% | 2% | 3% | 2% | 2% |

The majority of participants indicated they would spend under $10,000 for prevention and preparedness. Few individuals considered spending more than $100,000, e.g., a low of 10 individuals for landslides and a high of 20 individuals for severe weather. The expenditures are similar across all of the natural hazards, which implies that most individuals could be unfamiliar with the types of mitigation required for each of the hazards at the job choice location.

We use the round 4 survival rate to assess the influence of a hazard(s) spillover using the $AA$ based on experiment participants' information. We divide the $BR$ into 3 categories of financial burden for all 16 job profile IDs in Table 9. The categories are: Low – an individual retains 50% or more of their disposable income, Moderate – an individual retains some but less than 50% of disposable income, and High – an individual either breaks even or has a negative disposable income. Again, the bold numbers are for the leading jobs.

**Table 9 — Percent of financial burden by $FJC$ and job profile ID choice**

| | Final Job Choice Burden Level | | | | | |
|---|---|---|---|---|---|---|
| | **Minimum threshold** | | | **Maximum threshold** | | |
| **Job profile ID** | **Low** | **Moderat** | **High** | **Low** | **Moderat** | **High** |
| *All jobs* | *81.7%* | *5.1%* | *13.1%* | *48.9%* | *9.4%* | *41.7%* |
| 1 | 66.7% | 11.1% | 22.2% | 33.3% | 22.2% | 44.4% |
| 2 | 22.2% | 0.0% | 77.8% | 11.1% | 0.0% | 88.9% |
| **3** | **93.8%** | **2.1%** | **4.2%** | **56.3%** | **6.3%** | **37.5%** |
| **4** | **87.5%** | **1.8%** | **10.7%** | **48.2%** | **5.4%** | **46.4%** |
| 5 | 62.5% | 12.5% | 25.0% | 12.5% | 25.0% | 62.5% |
| 6 | 80.0% | 10.0% | 10.0% | 30.0% | 30.0% | 40.0% |
| **7** | **92.5%** | **2.5%** | **5.0%** | **70.0%** | **5.0%** | **25.0%** |
| 8 | 44.4% | 11.1% | 44.4% | 11.1% | 0.0% | 88.9% |
| 9 | 73.3% | 20.0% | 6.7% | 20.0% | 40.0% | 40.0% |
| **10** | **100.0%** | **0.0%** | **0.0%** | **60.5%** | **7.9%** | **31.6%** |
| 11 | 40.0% | 10.0% | 50.0% | 20.0% | 0.0% | 80.0% |
| 12 | 53.3% | 13.3% | 33.3% | 46.7% | 6.7% | 46.7% |
| **13** | **100.0%** | **0.0%** | **0.0%** | **66.7%** | **4.2%** | **29.2%** |
| 14 | 50.0% | 0.0% | 50.0% | 10.0% | 0.0% | 90.0% |
| 15 | 66.7% | 27.8% | 5.6% | 38.9% | 27.8% | 33.3% |
| **16** | **90.3%** | **0.0%** | **9.7%** | **67.7%** | **6.5%** | **25.8%** |

Inspection of Table 9 allows us to compare the average of all jobs to each of the job profile IDs to assess whether the financial burden of hazard prevention is above or below the baseline. Based on the survival in Figure 4, the table shows that preferred job profile IDs, such as 3, 4, 7 and 10, 13, and 16 are more likely to result in a low financial burden because they are in high salary locations. If we calculate the burden for job profile IDs based on the maximum prevention costs, only job profile IDs 3, 7, 10, 13, and 16 have greater than 50% of the individuals in the low burden category. On the other hand, job profile IDs that are less preferred and are lower salary include job profile IDs 1, 2, 5, 8, 11, 12, and 14, which have higher burden ratios than the baseline at both the minimum and maximum prevention cost levels. Matching the highest burden levels to job profile IDs 2, 8, 11, and 14 shows that these job offers are lower salary, generally have medium to higher crime rates (job profile ID 8 has a low to medium crime rate), and, for the most part, have lower exposure to natural hazard risks. Job profile IDs 2, 8, and 14 have very low to medium hazard exposure, yet have the highest financial burden exposure to natural hazards. This outcome may indicate an unequal distribution of income, or an overestimation by the participant on what they believe they can afford.

We combine salary income, cost of living and the minimum burden ratio to estimate the $AA$ in equation 13 for $FJC$ IDs. Assuming job seekers and screeners are cost minimizers, the $AA$ of natural hazard protection becomes:

$$(13)\quad AA = \mathcal{A}\left(\frac{Sal/CoL}{Min(BR)}\right)$$

$AA$ indicates if an individual would be capable of paying for natural hazard mitigation with the disposable income available from an $FJC$. Equation 13 provides a way to assess the cost-effectiveness of mitigation to manage natural hazard risks. Additionally, equation 13 is useful to classify the economic impact of hazard risks as they relate to whether an individual financial burden is low, moderate, or high in a $FJC$ location. In Figures 5a and 5b $Sal/CoL$ is plotted along the horizontal axis, and an individual's estimated $BR$ is plotted along the vertical axis. Salary level categories are shown as dashed lines on the graph. A point in the figure represents the anticipated affordability for an individual of their financial manageability for hazard protection. The figure illustrates a wide range of affordability and that high burden levels are found in all three salary categories. However, upon inspection of the graph, there is a concentration of higher burden levels at lower salary-to-cost of living ratios and a concentration of relatively easy manageability of hazard risk by individuals with lower hazard burden at higher salary-to-cost of living ratios. The value for $BR$ in the denominator is critical to the decision because it determines a self-protection limit. This limit is an individual's income left over after paying the prevention cost. $BR$ in Figures 5a and 5b is represented by 3 colors: low burden in blue, moderate burden in yellow, and high burden in red. The points on the graph take on a U shape that is similar to other models that involve tradeoffs between socioeconomic status and environmental disamenities (Kuznets 1955, Abdelfattah et al 2023, Botzen et al 2019).

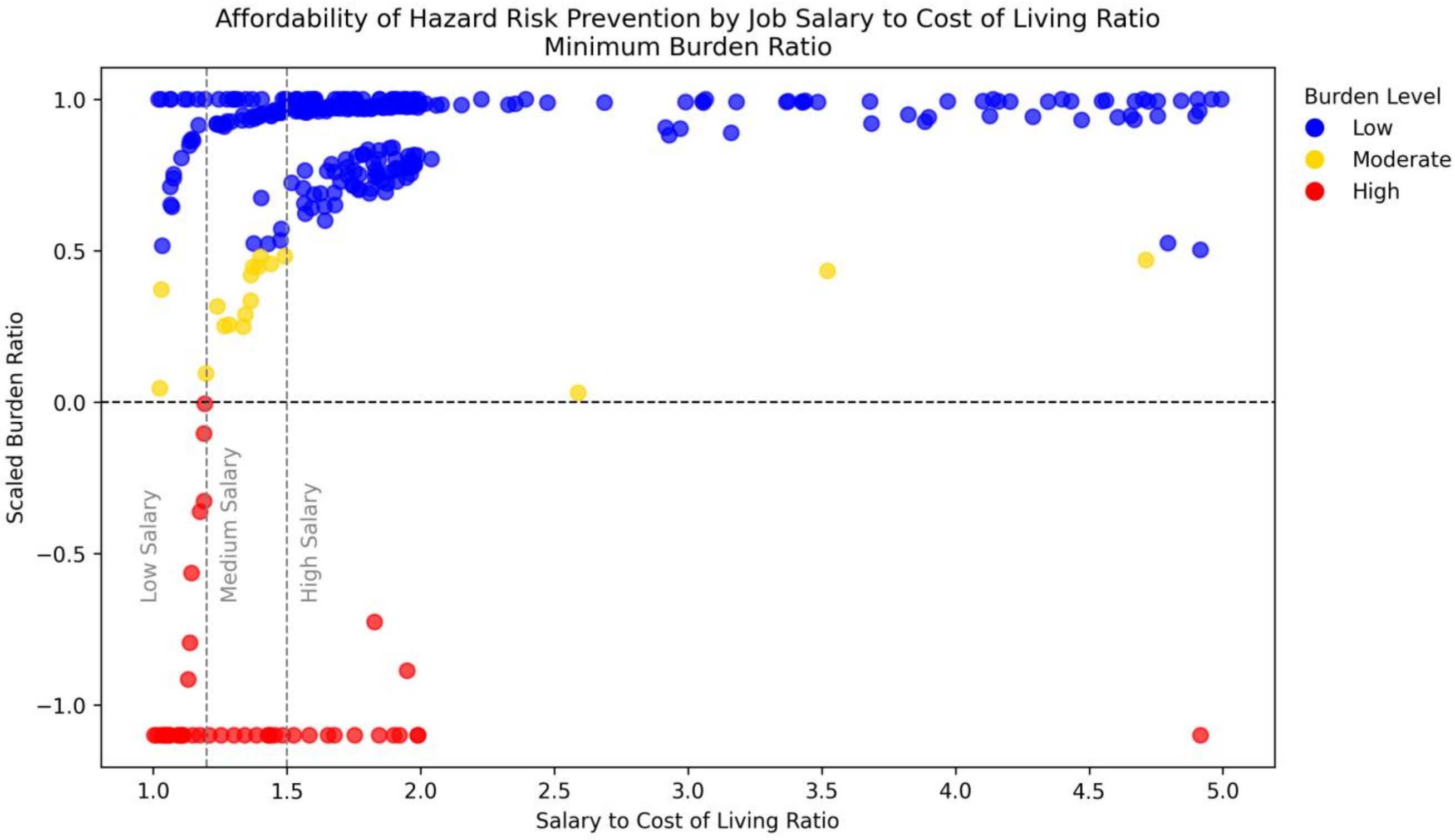

**Figure 5a. Anticipated Affordability of Natural Hazard Risk Minimum Burden Ratio by Salary / Cost of Living**

*Notes:* (Low – an individual retains 50% or more of their disposable income, Moderate – an individual retains some but less than 50% of disposable income, and High – an individual either breaks even or has a negative disposable income.)

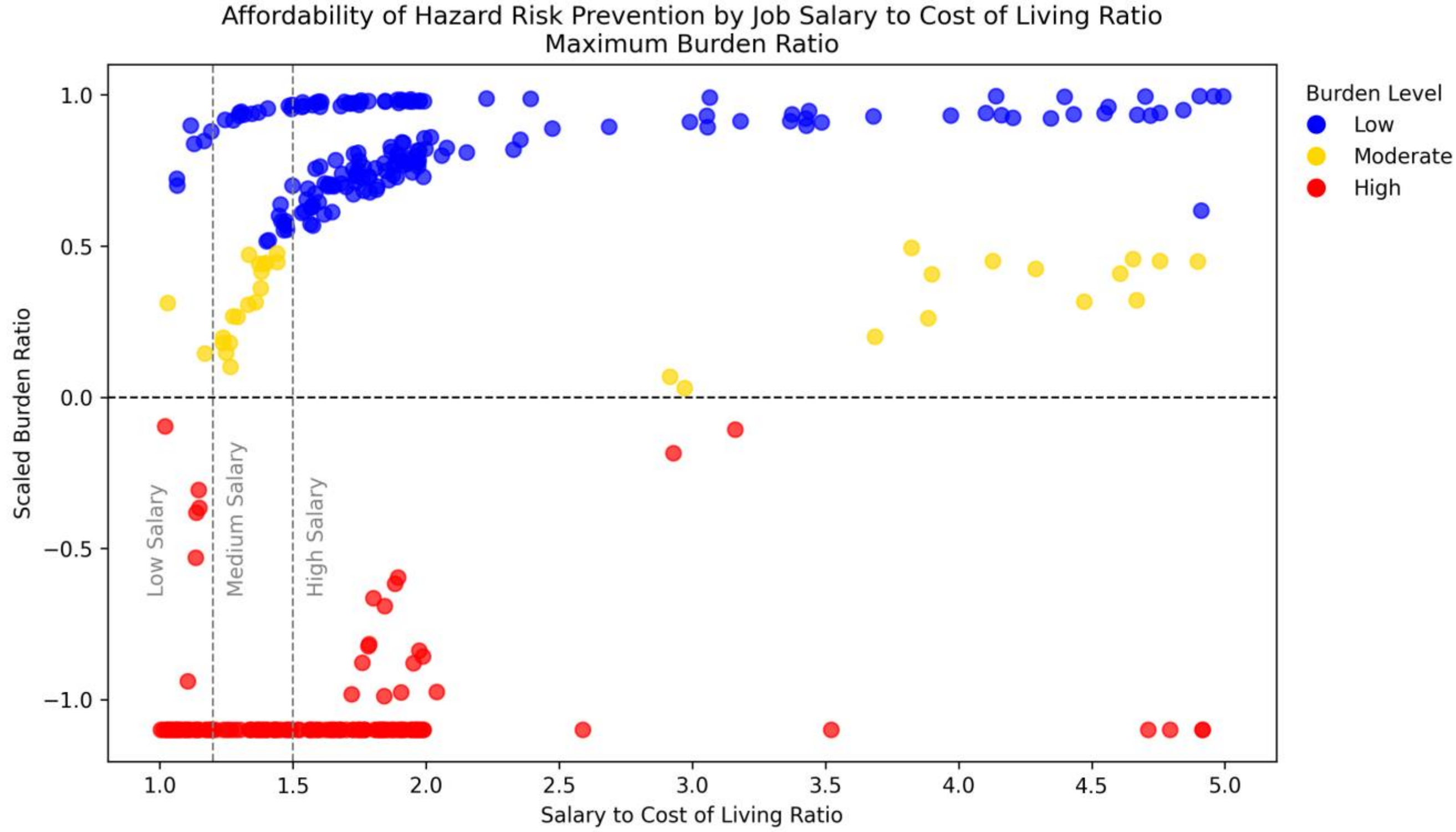

**Figure 5b. Anticipated Affordability of Natural Hazard Risk Maximum Burden Ratio by Salary / Cost of Living**

*Notes:* (Low – an individual retains 50% or more of their disposable income, Moderate – an individual retains some but less than 50% of disposable income, and High – an individual either breaks even or has a negative disposable income.)

For the minimum burden, the choice experiment resulted in 46 individuals with $FJC$s that fell below the breakeven point of $BR = 0$, 18 $FJC$s that were of moderate affordability, and 286 with a low $BR$. For maximum burden, the results showed 146 individuals with $FJC$s that fell below the breakeven point of $BR = 0$, 33 $FJC$s that were of moderate affordability, and 171 with a low $BR$. Figure 6 divides affordability for natural hazard risk protection into low, medium, and high salary categories. As expected, affordability is least likely for low salary individuals and unaffordability is unlikely to affect a high salary individual's loss prevention decision. Specifically, Figure 6 shows that 17% of low salaried individuals would be able to afford hazard protection even at the minimum $BR$, while 60% of medium salary individuals and 66% of high salary individuals could easily manage the cost of preparedness and prevention. At the breakeven point, or potentially unaffordable prevention cost, all three salary categories have significant percentages of respondents (Low: 33%; Medium: 25%; High: 29%).

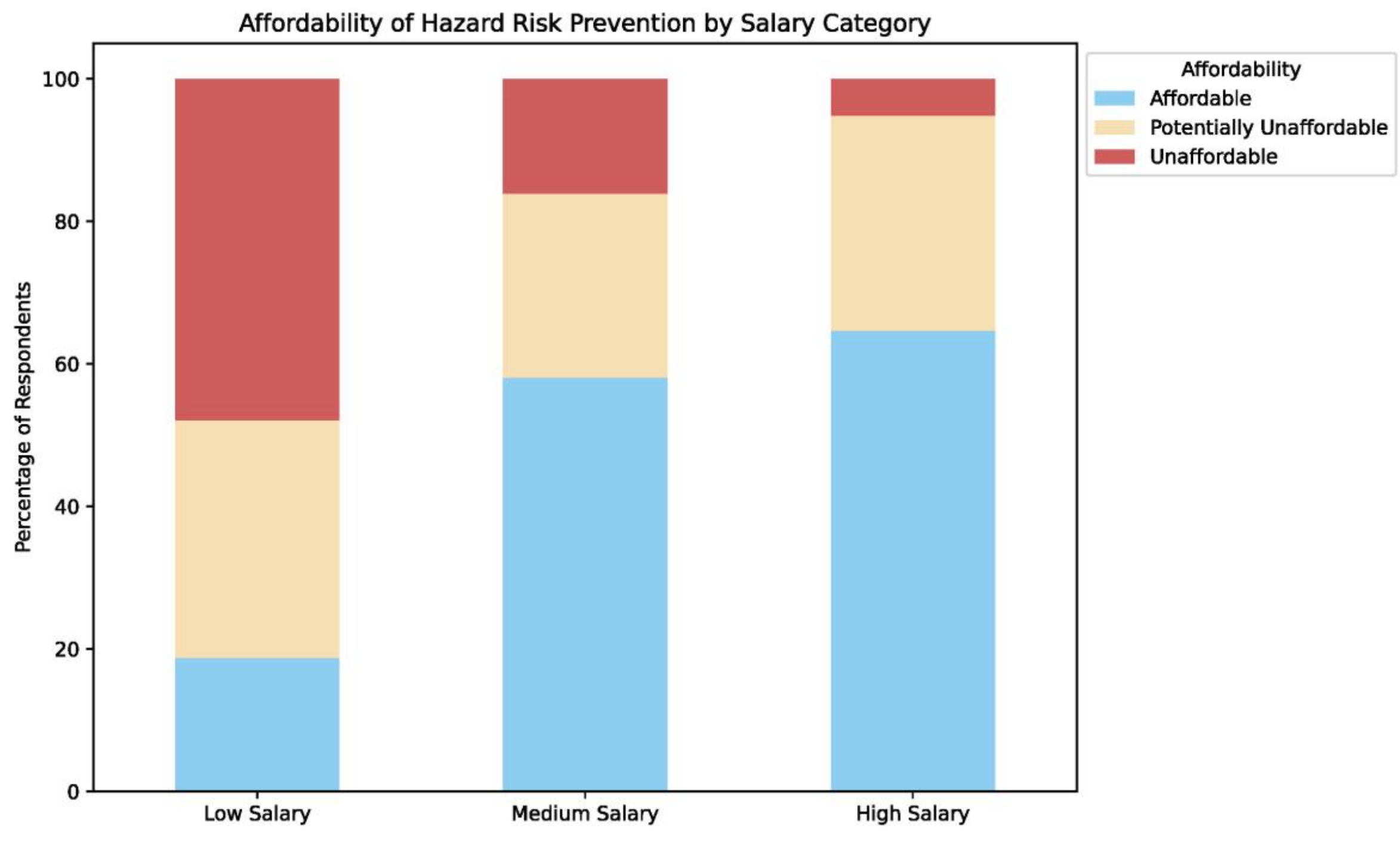

**Figure 6. Affordability of Hazard Risk Protection by Salary Category**

We selected a few salary and burden levels for specific individuals to reveal how $FJC$ job profile ID attribute level combinations are the inputs for identifying points in Figure 5a. One job profile ID's $FJC$ is a science-based offer for a Glaciology Meeting and Event Planner located in Fife, WA (near Seattle, WA) that has a high crime level, with natural hazards risks that are high for earthquakes, low for wildfires, medium for floods, medium for severe weather, very high for landslides, and very high for volcano risks. This entry has a Low Burden that yields an $AA$ of 0.498. For the same job profile ID, another individual chose a Research Associate position in Atlantic City, NJ that has a high crime level, with natural hazards risks that are low for earthquakes, medium for wildfires, very high for floods, high for severe weather, low for landslides, and none for volcano risks. This entry has a Low Burden that yields an $AA$ of 1.299. An Earth Science

Educational and Career Counselor in Mission Viejo, CA has a medium crime level, with natural hazards risks that are very high for earthquakes, high for wildfires, high for floods, medium for severe weather, medium for landslides, and none for volcano risks. This entry has a High Burden that yields an $AA$ of -84.608. And a job profile ID for a Geoscience Emergency Management Director in Des Arc, AR that has a medium crime level, with natural hazards risks that are low for earthquakes, very low for wildfires, low for floods, medium for severe weather, very low for landslides, and none for volcano risks. This entry has a High Burden that yields an $AA$ of -0.06. We highlight these choices to show the range of the variability among the inputs involved in a job search. It is clear that individuals simultaneously must weigh that variability in their search to be able to pick all the best attributes for an $FJC$. Therefore, selection is likely to be the outcome of a dynamic learning process. In the Appendix, we present several examples of $FJC$s and their associated crime levels and natural hazard category rankings to illustrate our approach. Each selection contains specific attributes about the job offer as identified from the 16 job profile IDs.

## 7. Discussion

Our results demonstrate the importance of relaxing the proportional hazards assumption in settings where preferences may evolve across time steps (Allison 1982). The ECH model reveals patterns that are masked under a PH specification and provides a flexible framework for incorporating time-varying effects into discrete choice environments (Han and Hausman 1990). More broadly, the paper illustrates how survival models can be used to represent multistage economic decisions, offering a bridge between two literatures (RUM and statistical hazard models) that are often treated separately. The Random Utility Model revealed that the timing of job elimination influences the conditional probability of job survival. The Kaplan – Meier estimation provides a conditional probability that a job remains under consideration through a given round and the Cox hazard ratio measures the relative effect of a covariate on the risk of elimination. The effect of the structural time-dependent term in equation 10 indicates that the influence of hazard risk becomes more pronounced as time passes and as salary increases through the learning process. We found that alternatives with lower utility are more likely to be eliminated earlier, while higher-utility alternatives persist through successive rounds and are more likely to be selected. The ECH model captures the possibility that the relative importance of specific attributes - particularly natural hazard risk - changes as individuals move from initial screening to final decision making. A sign flip in the $NHR$ effect occurs in the last round due to the covariate's effect not being constant through the experiment rounds. Instead, its effect depends on where the decision process is at a point in time. Thus, the $NHR$ effect is negative in early rounds of broad screening and in the final rounds it is positive for tradeoffs among strong candidates. We found that natural hazard risks may be tolerated and manageable for higher salary, positive location amenities, and lower crime levels. This is substantially meaningful in an adaptive context. In practical terms, this means that even though individuals are generally more likely to choose higher-paying jobs, some participants may still avoid high-hazard jobs despite the financial incentive. Interactions with crime levels and cost of living were not significant, suggesting that salary may more easily compensate for those factors in decision making.

In addition, results from the larger choice experiment study (Gonzales et al 2026) identified several salient points that relate to participants' decision-making processes. Income was generally rated more important for the $FJC$ than for the choice of the participant's current location of residence, while location was moderately important for both current location and final job choice. Notably, while hazard risk was one of the least important factors in final job choice and the choice of the current location of residence, this factor was rated slightly more important for final job choice than for choice of current location. In addition, the hazard risk level and participants' concern for hazards at their current location was generally higher across hazards than the actual level of impact experienced from those hazards. Additional insights from Gonzales et al 2025 indicated that hazard risk exposure at the current location of residence was greatest for severe weather, slides, and floods among participants, with most participants living in high to very high-risk areas for severe weather, moderate risk areas for slides, and 42% living in areas of moderate flood risk. However, with regard to impacts from hazards, only a third of participants reported medium severity impacts while most reported low or no impacts. In fact, severe weather had the highest levels of reported impacts (low to medium severity), followed by floods and wildfires (none to low severity). Severe weather also had the highest concern levels of any hazard. Taken together, these findings further support that financial considerations drive participants choices of $FJCs$, and while hazard risk is a concern, there is an underlying level of acceptance of some hazard risk in the choice of where to live.

This type of information enables public sector policymakers to obtain an ability to target areas for investment in loss avoidance with more precision to help prioritize public sector infrastructure decisions based on community financial vulnerability. Targeted hazard mitigation is cost effective as a policy initiative and as a risk management approach (Bernknopf and Amos 2014). The questionable rationality of ignoring natural hazard risk in a job search provides an example of the difficulty of combining experiential and analytical approaches in dealing with outcomes that change very slowly over time, are remote in time, and have been instinctual in nature.

The seek-and-screen approach provides a way to reduce the uncertainty concerning the natural hazard disamenity in selecting a job offer and to determine if the investment in private risk protection is expected to be affordable. Our results indicate that the participants in the adaptive risk experiment were concerned about natural hazards risks, i.e., natural hazard risk can influence choice at higher salary levels. Furthermore, they defer the use of this information until later rounds. That is, hazard risks are considered after salary and crime risk thresholds are met. It is clear that the effect of natural hazard information is statistically significant to managing natural hazard risk. Although natural hazard risks are alarming, they vary across location, recurrence, and severity and can be ignored due in part to the low spatial probability of events over a lifetime. Severe events can result in human and economic disasters that take place infrequently, but smaller, manageable risks recur episodically as hundreds of episodes with less consequences.

Based on the choice experiment, individuals can gain valuable insights into their hazard insurance choices and the need to invest in self-protection. As shown in Figures 5 and 6, for most employment choices, natural hazard risks would be manageable.

## 8. Conclusion

This paper develops and estimates a sequential model of job choice that integrates a Random Utility Model framework with a hazard-based representation of decision-making. By modeling job selection as a series of elimination events, the approach captures the dynamic nature of the seek and screen process and job acceptance. Our analysis provides a tractable link between discrete choice theory and survival analysis. The equivalence between the Cox partial likelihood and conditional logit probabilities over the risk set offers a coherent economic interpretation of the hazard model, in which job attributes influence the probability that an alternative is removed from consideration at each time step.

The empirical results highlight the distinct roles of economic and environmental job attributes. Salary relative to cost of living and crime rate exhibit stable and economically intuitive effects throughout the decision process, reinforcing their importance as core determinants of job desirability. In contrast, natural hazard risk displays a time-varying and non-monotonic influence: it acts as a screening criterion in early rounds but becomes less detrimental—and ultimately favorable—among the final set of alternatives. This finding suggests that individuals evaluate environmental risks differently depending on the stage of the decision process and the composition of the remaining choice set.

We show that there is ample evidence that individuals are aware of hazard risks and use the information to aid in a job offer decision at later stages in the decision-making process. Salary is the most controlling factor in final job selection, with higher-paying jobs significantly more likely to be chosen. We can confirm that compensation and the effect of cumulative hazard risk do play a dominant role in shaping a job decision. This result implies that concerns about safety persist even in the face of financial rewards. Since the final job choice represents the set of surviving preferred attributes of a specific job offer at a specific time, it provides us with the impact of the factors that can be valuable to individuals and economic sectors.

Several avenues for future research remain. First, incorporating individual-level heterogeneity and unobserved preference variation would allow for richer behavioral insights. Second, extending the framework to observational data would enable the evaluation of policy-relevant questions related to labor mobility and environmental risk exposure. Finally, integrating dynamic search behavior and expectations about future opportunities could further enhance the realism of the model. Overall, the findings underscore the value of modeling job choice as a dynamic process and highlight the complex role of environmental risk in shaping labor market outcomes.

## References

Abdelfattah, Y., Shireen AlAzzawi, Nada Rostom, and Heba Abdelkader. 2023. "The Inequality, Economic Growth, Climate Change and Natural Disasters Nexus: Empirical Evidence." Economic Research Forum 29th Annual Conference, Cairo, Egypt. https://erf.org.eg/app/uploads/2023/04/1681210943_430_1283833_131erf29ac_clm_abdelfattah_alazzawi.pdf.

Allison, Paul. 1982. "Discrete-Time Methods for the Analysis of Event Histories." *Sociological Methodology* 13: 61-98.

Balboni, Claire. 2025. "In Harm's Way? Infrastructure Investments and the Persistence of Coastal Cities." *American Economic Review* 115 (1): 77-116.

Bayer, Patrick, Nathaniel Keohane, and Christopher Timmins. 2009. "Migration and hedonic valuation: The case of Air Quality." *Journal of Environmental Economics and Management* 58 (1): 1-14.

Bernknopf, Richard, and Paul Amos. 2014. “Measuring Earthquake Risk Concentration for Hazard Mitigation.” *Natural Hazards: Journal of the International Society for the Prevention and Mitigation of Natural Hazards* 74 (3): 2163-2192.

Bernknopf, Richard, David Brookshire, and Philip Ganderton. 2003. *The Role of Geoscience Information in Reducing Catastrophic Loss Using a Web-based Economics Experiment*. USGS Professional Paper 1683, U.S. Geological Survey.

Bishop, Kelly. 2012. *A dynamic model of location choice and hedonic valuation*. Mimeo, Washington University in St. Louis, https://scholar.google.com/scholar?q=A+Dynamic+Model+of+Location+Choice%0D%0Aand+Hedonic+Valuation&hl=en&as_sdt=0&as_vis=1&oi=scholart.

Brookshire, David, Mark Thayer, John Tschirhart, and William Schulze. 1985. “A Test of the Expected Utility Model: Evidence from Earthquake Risks” *Journal of Political Economy* 93 (2): 369-389.

Botzen, W.J., Oliver Deschenes, and Mark Sanders. 2019. “The Economic Impacts of Natural Disasters: A Review of Models and Empirical Studies.” *Review of Environmental Economics and Policy* 13 (2): 167-188. doi: 10.1093/reep/rez004.

California Department of Forestry and Fire Protection. (2025). 2025 Incident Archive. https://www.fire.ca.gov/incidents/2025.

Carstensen, Christian. (2020). Three essays on housing markets: Price dispersion, dynamic location choices and family investments, PhD Series, No. 212, University of Copenhagen, Department of Economics, Copenhagen. https://hdl.handle.net/10419/240560.

Economou, Theodorous, David Stephenson, J.C. Rougier, Robert Neal, and Ken Mylne. 2016. "On the Use of Bayesian Decision Theory for Issuing Natural Hazard Warnings." *Proceedings of the Royal Society A* 472:20160295, doi:10.1098/rspa.2016.0295.

Ehrlich, Isaac, and Gary Becker, G. 1972. "Market Insurance, Self-Insurance, and Self-Protection." *Journal of Political Economy* 80 (4): 623-648.

Fajgelbaum, Pablo, and Cecile Gaubert. 2020. "Optimal spatial policies, geography and sorting." *Quarterly Journal of Economics* 135 (2): 959-1036. doi: 10.1093/qje/qjaa001

Federal Bureau of Investigation. [FBI]. 2022. *National Incident-Based Reporting System (NIBRS) Tables: State Tables, Offenses by Agency: United States Offense Type by Agency, 2022: population, total offenses, crimes against persons, crimes against property, and crimes against society.* https://cde.ucr.cjis.gov/LATEST/webapp/#/pages/downloads

Federal Emergency Management Agency. [FEMA]. 2025. *National Risk Index Data: Methodology and Hazards Overview*. https://www.fema.gov/sites/default/files/documents/fema_national-risk-index_methodology-hazards-overview.pdf

Gati, Itamar, and Itay Asher. 2001. "The PIC Model for Career Making: Prescreening, In-Depth Exploration, and Choice." Leong, Frederick, and Azy Barak, (eds.) *Contemporary Models in Vocational Psychology: A Volume in Honor of Samuel H. Osipow*. Lawrence Erlbsky, aum Associates, Publishers, Mahwah, NJ: 7-54.

Gati, Itamar, Nimrod Levin, and Shiri Landman-Tal. 2019. "Decision-Making Models and Career Guidance." Athanasou, J., and Perera, H. (eds.), *International Handbook of Career Guidance*. Springer Nature: 115-145. https://doi.org/10.1007/978-3-030-25153-6_6

Gourevitch, Jesse, Carolyn Kousky, Yanjin Liao, Christoph Nolte, Adam Pollack, Jeremy Porter, and Joakim Weill. 2023. Unpriced climate risk and the potential consequences of overvaluation in US housing markets. Nature Climate Change, 13, 250-257. https://doi.org/10.1038/s41558-023-01594-8.

Grossi, Patricia, and Howard Kunreuther, H. 2005. *Catastrophe Modeling: A New Approach to Managing Risk*. Springer Science + Business Media.

Gonzales, Leila, Christopher Keane, and Richard Bernknopf. 2025. "Understanding how geoscientists prioritize natural hazard risk in their decision-making." https://grande.americangeosciences.org/data/hazard-game/

Gonzales, Leila, Christopher Keane, and Richard Bernknopf. 2026. "The gap between attitudes and action within the US geoscience community's response to natural hazards." *Geoscience Communication* 9 (1): 35-67. doi: 10.5194/gc-9-35-2026.

H. John Heinz III Center for Science, Economics and the Environment. 2000. *The Hidden Costs of Coastal Hazards: Implications for Risk Assessment and Mitigation.* Island Press, Washington, D.C.

Han, Aaron, and Jerry Hausman. 1990. "Flexible Parametric Estimation of Duration and Competing Risk Models." *Journal of Applied Econometrics* 5 (1): 1–28.

Howard, G., and Liebersohn, J. (2025). How Regional Inequality and Migration Drive Housing Prices and Rents, Journal of Economic Perspectives, v.39, p.3-26.

Husted, T., and Nickerson, D. (2019). Disaster Risk, Moral Hazard, and Public Policy, Oxford Research Encyclopedia, DOI: 10.1093/acrefore/9780199389407.013.195.

Kellenberg, D., and A. Mobarak. (2008). Coes rising income increase or decrease damage risk from natural disasters? Journal of Urban Econoimics, v.63, p.788-802.

Kennan, J., and Walker, J. (2011). The Effect of Expected Income on Individual Migration Decisions, Econometrica, v.79, p.211-251.

Kiefer, N. (1988). Economic Duration Data and Hazard Functions, Journal of Economic Literature, v.26, p.646-679.

Kleinbaum, D. (1996). Survival Analysis: A Self-learning Text, Springer-Verlag, New York, NY.

Kousky, C., E. Luttmer, and R. Zeckhauser. (2006). Private investment and government protection, Journal of Risk and Uncertainty, v.33, p.73-100. https://doi.org/10.1007/s11166-006-0172-y.

Kunreuther, H., and Pauly, M. (2004). Neglecting Disaster: Why Don't People Insure Against Large Losses? Journal of Risk and Uncertainty, v.28, p.5-21.

Kuznets, S. (1955). Economic Growth and Income Inequality, American Economic Review, v.45, p.1-28.

Lancaster, T. (1997). The Econometric Analysis of Transition Data, Cambridge University Press, Cambridge, UK.

Landry, C., P. Hindsley, O. Bin, J. Kruse, J. Whitehead, and K. Wilson. (2011). Weathering the Storm: Measuring Household Willingness-to-Pay for Risk-Reduction in Post-Katrina New Orleans, Southern Economic Journal, v.77, p.991-1013.

Lewis, T., and D. Nickerson. (1988). Self-Insurance against Natural Disasters, Joural of Environmental Economics and Management, v.16, p.209-223.

Los Angeles Times. (2026). L.A.'s double disaster left thousands of scars, and the healing will take years. https://www.latimes.com/after-the-fires.

Louviere, J. (2001). Choice Experiments: An Overview of Concepts and Issues, in Bennett, N., and Blamey, R, ed., The Choice Modelling Approach to Environmental Valuation, Edward Elger Publishing, Inc., Northampton, MA., p.13-36.

McFadden, D. (1974). Conditional Logit Analysis of Qualitative Choice Behavior, in Frontiers in Econometrics, Paul Zarembka (ed.), Academic Press, p. 105–142.

Molloy, R., Smith, C., and Wozniak, A. (2011). Internal Migration in the United States, Journal of Economic Literature, v.25, p.173–196.

Ratcliffe, C., Congdon, W., Stanczyk, A., Teles, D., Martin, C., and Kotapati, B. (2019). Insult to Injury: Natural Disasters and Residents' Financial Health, Urban Institute, Washington, DC.

Restrepo, P. (2017). The United States National Weather Service Real-Time Flood Forecasting, Oxford Research Encyclopedia Hazard Science, Oxford University Press. DOI: 10.1093/acrefore/9780199389407.013.128.

Shavell, S. (2014). A General Rationale for a Governmental Role in the Relief of Large Risks, Harvard John M. Olin Center for Law, Economics, and Business, Discussion Paper No. 768, May 2014.

Slovic, P., Finucane, M., Peters, E., and MacGregor, D. (2004). Risk as Analysis and Risk as Feelings: Some Thoughts about Affect, Reason, Risk, and Rationality, Risk Analysis, v.24, p.311-322.

Viscousi, W.K. (1992). Fatal Tradeoffs Public and Private Responsibilities for Risk, Oxford University Press, New York, NY.

Viscousi, W.K., and Zeckhauser, R. (2006). National survey evidence on disasters and relief: Risk beliefs, self-interest, and compassion, Journal of Risk and Uncertainty, v.33, p.13-36. doi: 10.1007/s11166-006-0169-6.

Yezer, A. (2010). Expectations and unexpected consequences of public policy toward natural and man-made disasters, The Economics of Natural and Unnatural Disasters, v.3, 39-64.

## Supplementary Appendix A: Selections of FJCs and their specific amenity attributes of the job offer as identified from the job profile IDs, and Anticipated Affordability

| Job ID | PID[1] | Job Title | Location | Crime Risk | Natural Hazard Risk Level[2] | | | | | | Anticipated Affordability[3] |
|---|---|---|---|---|---|---|---|---|---|---|---|
| | | | | | EQ | WF | FL | SW | LS | VO | |
| **1** | 11 | Glaciology Meeting & Event Planner | Fife WA | H | H | L | M | M | VH | VH | 0.498 (LB) |
| **1** | 18 | Research Associate | Atlantic City NJ | H | L | M | VH | H | L | None | 1.299 (LB) |
| **2** | 17 | Postal Service Clerk | Watertown NY | H | L | VL | L | H | M | None | 1.021 (LB) |
| **3** | 7 | Acoustic Physicist | Danville NH | L | L | VL | M | H | M | None | 10.047 (MB) |
| **3** | 12 | Hydrology Logistics Engineer | Lockhart AL | L | VL | VL | L | H | M | None | 2.626 (LB) |
| **4** | 4 | Geomorphology Web Administrator | Lincoln MA | L | M | L | H | H | M | None | -2.518 (HB) |
| **5** | 2 | Earth Science Educational & Career Counselor | Mission Viejo CA | M | VH | H | H | M | M | None | -84.608 (HB) |
| **5** | 9 | Geomorphology Specialist | Stallings NC | M | L | VL | L | VH | L | None | 2.772 (MB) |
| **8** | 5 | Geoscience Emergency Management Director | Des Arc AR | M | L | VL | L | M | VL | None | -0.06 (HB) |
| **11** | 6 | Equipment Operator for Paving, Surfacing, & Tamping | Spanish Fork UT | M | H | H | M | H | H | None | -11.442 (HB) |
| **11** | 15 | Mixing & Blending Machine Setter, Operator, & Tender | Baytown TX | MH | L | M | VH | VH | M | None | 1.635 (LB) |
| **12** | 1 | Climatology Researcher | Grants Pass OR | MH | M | H | H | M | VH | None | -33.043 (HB) |
| **12** | 3 | Geomorphology Specialist | Highland Township MI | M | L | L | M | VH | M | None | -0.123 (HB) |
| **13** | 13 | Atmospheric Science Specialist | North Providence RI | M | L | VL | M | H | M | None | 4.785 (LB) |
| **15** | 8 | Photonics Technician | Twin Falls ID | MH | L | M | L | H | L | None | 4.075 (MB) |
| **16** | 10 | Atmospheric Science Compliance Manager | Longmont CO | MH | L | M | H | H | H | None | 9.133 (LB) |
| **16** | 14 | Geoscience Practitioner | Bristol Bay Borough AK | MH | VL | VL | None | VL | None | VL | 1.731 (LB) |
| **16** | 16 | Oceanographer | Buffalo Grove IL | M | L | VL | M | VH | M | None | 4.958 (LB) |

[1]PID = Participant ID. Randomized value assigned to final job choice details for participant final job choices.

[2]LB = low burden, MB = moderate burden, HB = high burden

[3]Natural Hazard Risk Level: EQ = Earthquakes; WF = Wildfires; FL = Floods; SW = Severe Weather; LS = Landslides; VO = Volcanoes